\documentclass[a4paper,12pt]{article}%
\usepackage{amssymb}
\usepackage{graphicx}
\usepackage{amsbsy}
\usepackage{amsmath}
\usepackage{cite}
\usepackage{slashed}
\usepackage{amsfonts}
\usepackage{hyperref}
\usepackage{array}
\usepackage{caption}
\usepackage{multirow}
\usepackage[compat=1.0.0]{tikz-feynman}
\hypersetup{
colorlinks   = true,
linkcolor    = blue,
citecolor    = blue,
}

\begin{document}
\thispagestyle{empty} \setcounter{page}{0} \begin{flushright} February 2025\\
\end{flushright}

\vskip3.4 true cm

\begin{center}
{\huge Dark higher-form fields and triangle anomalies}\\[1.9cm]

\textsc{Cypris Plantier}$^{1}$\textsc{, Christopher Smith}$^{2}$\vspace
{0.5cm}\\[9pt]\smallskip{\small \textsl{\textit{Laboratoire de Physique
Subatomique et de Cosmologie, }}}\linebreak%
{\small \textsl{\textit{Universit\'{e} Grenoble-Alpes, CNRS/IN2P3, Grenoble
INP, 38000 Grenoble, France}.}} \\[1.9cm]\textbf{Abstract}\smallskip
\end{center}

\begin{quote}
\noindent
Light scalar and vector particles admit non-trivial descriptions in terms of anti-symmetric higher-rank tensor fields. Far from mere rewritings, these provide compelling alternative frameworks, leading to immediate phenomenological applications. In this paper, we extend the playground to include the contributions of fermionic triangle loops, and use these results to compare the standard and higher-form realizations for two phenomenological processes: the pair production of two dark particles and the decay of a dark particle in two photons. Though they all do show some dependencies on the chosen realizations for the spin-zero or the spin-one dark states, we find that the two-photon mode is particularly sensitive and could actually be our prime window into the true nature of the dark field.

\let\thefootnote\relax\footnotetext{\newline$^{1}\;$%
cplantier@lpsc.in2p3.fr\newline$^{2}$~chsmith@lpsc.in2p3.fr}
\end{quote}

\newpage

\setcounter{tocdepth}{2}%

\section*{Introduction}

Understanding the exact nature of dark matter (DM) remains one of the most important and compelling problem in particle physics. Although astrophysical and cosmological evidences leave little room for doubt regarding its existence, the immense set of theoretical possibilities coupled to the absence of information regarding the energy-domain to scrutinize tends to make this research a long-term effort. Among all the candidates, light, weakly-interacting spin-zero (like axions and ALPs) and spin-one (often referred to as dark photons \cite{Holdom:1985ag,Miller:2021ycl}) particles have been around for a long time now, but it is the absence of supersymmetric signatures at the LHC that truly brought these models to the forefront with the axion as the driving force (for recent reviews, see, e.g. \cite{DiLuzio:2020wdo,Smith:2024uer}).

If it is trivially possible to embody a spin-zero particle in a scalar field and a spin-one 
particle in a vector field, it is also possible to adopt representations based on the so-called higher-form fields. Those are fundamental, antisymmetric tensor fields that possess more than one Lorentz indices. Specifically, it is well-known that a massive spin-one state can be represented by an antisymmetric rank-2 tensor field $B^{\mu\nu}$, known as Kalb-Ramond field \cite{Kalb:1974yc}, whereas a massive spin-zero state can be embedded in an antisymmetric rank-3 tensor field $C^{\mu\nu\rho}$ \cite{Curtright:1980yj,Curtright:1980yk}, that we simply refer to as "3-form". These forms originally emerged in the context of string theory (for details, see e.g. \cite{Polchinski:1998rr,Svrcek:2006yi}) and have been mostly constrained since then to the domain of formal theory. The reason why these forms have been historically disregarded from a phenomenological point-of-view relies on the existence of algebraic prescriptions relating the standard and higher-forms descriptions of a same state: the so-called \textit{dualities} \cite{Polchinski:1998rr,Hjelmeland:1997eg}. The most famous examples are the electromagnetic duality and the massless scalar - 2-form duality (extensively used in the context of axion physics). Due to this, the standard and higher-forms descriptions are often mistaken as phenomenologically equivalent.

However, the recent renewal in the interest for higher-forms and dualities \cite{Malta:2025ydq,Burgess:2025geh} has enlightened a certain number of important features (for a complete phenomenological review, see \cite{Plantier:2025hcm}). Among them, the fact that duality reduces to a strict equivalence only in the context of free theories. Whenever we introduce couplings to external fields we jeopardize this equivalence and at the end we obtain two distinct theories with distinct signatures. Specifically, it appears that the dualization process does not commutes with the mass power-counting at the level of the Lagrangian. As a consequence, two dual forms will have different dominant couplings with standard model (SM) fields. Considering a fermionic current, for example, a massive scalar field $\phi$ couples to it through $\bar{\psi}\psi\phi$ at the renormalizable level, whereas its dual partner, the massive 3-form $C^{\mu\nu\rho}$, couples through the operator $C_{\mu\nu\rho}\varepsilon^{\mu\nu\rho\sigma}\bar{\psi}\gamma_{\sigma}\psi$.

This last argument fully motivates the present paper: contrary to a vector field that couples vectorially or axially to the SM fermions, a 2-form field couples (pseudo)tensorially at the renormalizable level. If we want to mediate a process involving at least one massive dark spin-one particle through a fermionic triangle loop, we therefore expect two distinct contributions in the vector and 2-form case. Same holds for a process involving a spin-zero particle, due to the peculiar nature of the 3-form - fermions coupling. This will drive us to consider and calculate triangle diagrams that were previously considered irrelevant: indeed, triangles involving tensorial or antisymmetric vertices are systematically highly-suppressed in the SM, whereas they are dominant in a higher-form description of dark matter. The interest is also theoretical, as some of these triangles are \textit{a priori} susceptible to exacerbate anomalies, and specifically we will see in the following that dealing with the 3-form requires a careful and peculiar treatment of the ABJ anomaly~\cite{Adler:1969gk,Bell:1969ts}.

The paper is organized as follows: In a first section, we examine all the triangle diagrams that are susceptible to contribute in the later-considered processes. This will lead us to review in details the well-known axial-vector-vector ($AVV$) triangle, before diving into the treatment of triangles containing at least one (pseudo)tensorial vertex. 
The second section focuses on two simple phenomenological applications: the decay of a Standard Model scalar or pseudoscalar in two dark fields and the decay of a dark field in two photons.

\section{Triangle diagrams}

Higher-form fields open the gate to a great diversity of possible couplings with SM fields ~\cite{Plantier:2025hcm}. Focusing on the renormalizable couplings that each of these fields realizes with the SM fermions, we end up with a host of possible fermionic triangles to consider. Therefore, our goal will be to calculate all the relevant triangles explicitly before coupling them to external SM bosons or dark higher-form fields.

Let us start by labelling all the vertices we will encounter in order to cover all possible triangle amplitudes in a more generic way:
\begin{subequations}
\begin{align}
     \text{Scalar coupling:}\;\;\;\;\;\;\;\;\;\;\;\;\;\; \bar{\psi}\psi &\longrightarrow (S)\;,\\
     \text{Pseudoscalar coupling:}\;\;\;\;\;\;\;\;\;\; \bar{\psi}\gamma^5\psi &\longrightarrow (P)\;,\\
    \text{Vector coupling:}\;\;\;\;\;\;\;\;\;\; \bar{\psi}\gamma^{\mu}\psi &\longrightarrow (V)\;,\\
    \text{Axial coupling:}\;\;\;\;\;\; \bar{\psi}\gamma^{\mu}\gamma^5\psi &\longrightarrow (A)\;,\\
    \text{Tensorial coupling:}\;\;\;\;\;\;\;\; \bar{\psi}\sigma^{\mu\nu}\psi &\longrightarrow (T)\;,\\
    \text{Pseudotensorial coupling:}\;\;\;\; \bar{\psi}\sigma^{\mu\nu}\gamma^5\psi &\longrightarrow (\tilde{T})\;,
\end{align}
\end{subequations}
with $\sigma^{\mu\nu}=i[\gamma^{\mu},\gamma^{\nu}]/2$. In the first subsection, to set the stage, we will review in some details the well-known $AVV$ triangle calculation, along with those involving $S$ and $P$ vertices. Then, in the second subsection, we will consider the triangle diagrams involving one or more tensorial or pseudotensorial vertices. Those will play an important role in our phenomenological analyses because the leading fermionic couplings of a 2-form field are of that kind.

\subsection{The $AVV$ triangle}
Ultimately, one of our objectives is to compare the way 1-forms and 3-forms couple to SM gauge bosons through triangle diagrams. Focusing on renormalizable couplings, we will see that both the 1-form and 3-form processes rely on the well-known $AVV$ triangle~\cite{Adler:1969gk,Bell:1969ts}.

Although its amplitude and anomalous nature are well-known and well-described in the literature (for a detailed treatment, see e.g. Ref.~\cite{Weinberg:1996kr}), it is most of the time treated at the level of its divergences only. The reason is that, whether in the frame of chiral theory, for example to understand the $\pi^0$ decay~\cite{Bell:1969ts}, or in the frame or axion dynamics, the considered fields only couple to the divergence of the axial current. As a result, phenomenologically, only the divergence of the triangle amplitude is needed, and not the amplitude itself. Yet, as we will see, the 3-form field couples to the $AVV$ triangle directly, not to its divergence, so we need to characterize how the anomaly manifests itself at that level. Further, while assigning the anomaly to any of the current follows a well-established procedure at the divergence level, e.g. via a choice of momentum routing~\cite{Weinberg:1996kr}, it is less clear \textit{a priori} how to proceed at the amplitude level.
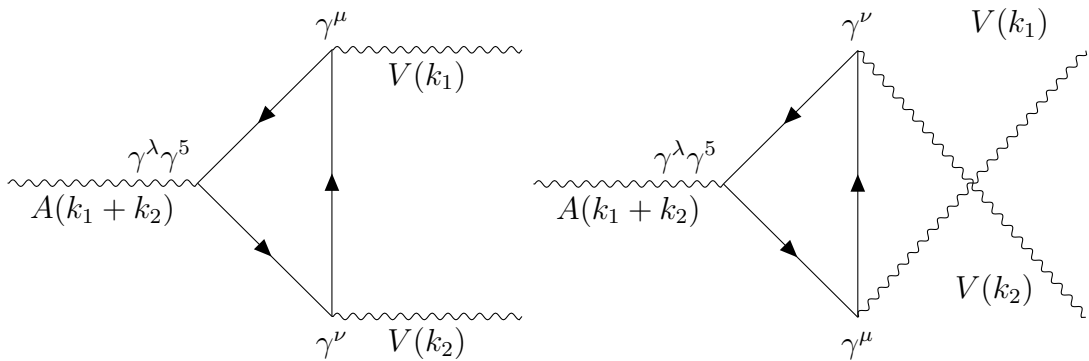
\begin{figure}[t]
\begin{tikzpicture}[scale=2]
 
  \begin{feynman}
    \vertex (a);
    \vertex [right=2cm of a,label=$\gamma^{\lambda}\gamma^5$](aa);
    \vertex [right=2.5cm of a] (b);
    \vertex [ below right=2.5cm of b] (c);
    \vertex [ below=0.7cm of c,label=$\gamma^{\nu}$] (cc);
    \vertex [ above right=2.5cm of b] (d);
    \vertex [left=0.0cm of d,label=$\gamma^{\mu}$](dd);
    \vertex [right=2.5cm of c] (e);
    \vertex [right=2.5cm of d] (f);

    \diagram* {
      (a) -- [photon,edge label'=\(A(k_1+k_2)\)] (b),
      (b) -- [fermion] (c),
      (c) -- [fermion] (d),
      (d) -- [fermion] (b),
      (d) -- [photon, edge label'=\(V(k_1)\)] (f),
      (c) -- [photon,edge label'=\(V(k_2)\)] (e),

    };
  \end{feynman}
\end{tikzpicture}
\begin{tikzpicture}[scale=2]
 
  \begin{feynman}
    \vertex (a);
      \vertex [right=2 cm of a,label=$\gamma^{\lambda}\gamma^5$](aa);
    \vertex [right=2.5cm of a] (b);
    \vertex [ below right=2.5cm of b] (c);
    \vertex [ below=0.7cm of c,label=$\gamma^{\mu}$] (cc);
    \vertex [ above right=2.5cm of b] (d);
    \vertex [left=0cm of d,label=$\gamma^{\nu}$](dd);
    \vertex [right=3cm of c] (e);
    \vertex [right=3cm of d] (f);
    \vertex [left=1.2cm of e, label=\(V(k_2)\)] (ee);
    \vertex [below= 0.5cm, left=1cm of f, label=\(V(k_1)\)] (ff);

    \diagram* {
      (a) -- [photon,edge label'=\(A(k_1+k_2)\)] (b),
      (b) -- [fermion] (c),
      (c) -- [fermion] (d),
      (d) -- [fermion] (b),
      (d) -- [photon] (e),
      (c) -- [photon] (f),

    };
  \end{feynman}
\end{tikzpicture}
\caption{Direct and crossed $AVV$ triangle diagrams}
\end{figure}

First, let us recall the expression of the $AVV$ triangle amplitude that takes in account both the direct and the crossed diagrams (see Fig.~1):
\begin{equation}
    T_{AVV}^{\lambda\mu\nu}=i\int\frac{d^4q}{(2\pi)^4}\text{Tr}\left[S(q-k_1)\gamma^{\lambda}\gamma^{5}S(q)\gamma^{\mu}S(q+k_2)\gamma^{\nu}\right]+\{\mu\leftrightarrow\nu,k_1\leftrightarrow k_2\}\;.
\end{equation}
With $S(k)^{-1}=\slashed{k}-m$, $m$ the mass of the fermion circulating in the loop, and $k_{1,2}$ the outgoing momenta (we will keep these labelling conventions in the following). At this stage, we are allowed to move the $\gamma^5$ matrix in the Dirac trace since $\{\gamma^{\mu},\gamma^5\}=0$ in four dimensions, so $\gamma^5S(k)=S(-k)\gamma^5$. But, even if ultimately finite, the loop integration formally diverges and needs to be regulated. When adopting dimensional regularization (DR), $\gamma^5$ ceases to anticommute, which means that the final result depends on where $\gamma^5$ was left before regularization. To account for this ambiguity, we follow Ref.~\cite{Elias:1982ea} (see also Ref.~\cite{Quevillon:2021sfz}) and introduce free $a,b$ parameters as 
\begin{align}
   \operatorname{Tr}\left[S(q-k_1)\gamma^{\lambda}\gamma^{5}S(q)\gamma^{\mu}S(q+k_2)\gamma^{\nu}\right]\rightarrow \; & (1-a-b)\operatorname{Tr}\left[S(q-k_1)\gamma^{\lambda}\gamma^{5}S(q)\gamma^{\mu}S(q+k_2)\gamma^{\nu}\right]\nonumber\\
   & -b\operatorname{Tr}\left[S(q-k_1)\gamma^{\lambda}S(-q)\gamma^{\mu}\gamma^5S(q+k_2)\gamma^{\nu}\right]\nonumber\\
   & +a\operatorname{Tr}\left[S(q-k_1)\gamma^{\lambda}S(-q)\gamma^{\mu}S(-q-k_2)\gamma^{\nu}\gamma^5\right],
\end{align}
with the understanding that these Dirac traces are to be evaluated in the BMHV scheme \cite{Breitenlohner:1977hr}. These $a,b$ parameters are strictly equivalent to those introduced to keep track of the momentum-routing ambiguity when using a cutoff regularization \cite{Weinberg:1996kr}. 

At this step, it is possible to perform a straightforward calculation using Passarino-Veltman reduction \cite{Passarino:1978jh}. To see exactly how the anomalous part contributes, we start as usual by considering the three divergences of the $AVV$ amplitude, that are the Ward identities of the underlying currents coupled to the triangle:
\begin{subequations}
\begin{align}
    -ik_{1\mu}T_{AVV}^{\lambda\mu\nu}&=\frac{b}{4\pi^2}\varepsilon^{\nu\lambda k_1 k_2}\ ,\\
    -ik_{2\nu}T_{AVV}^{\lambda\mu\nu}&=\frac{a}{4\pi^2}\varepsilon^{\lambda\mu k_1 k_2}\ ,\\
     i(k_{1}+k_{2})_{\lambda}T_{AVV}^{\lambda\mu\nu}&=\frac{m^2}{\pi^2}\text{C}_0(m^2)\varepsilon^{\mu\nu k_1 k_2}-\frac{(a+b-2)}{4\pi^2}\varepsilon^{\mu\nu k_1 k_2}\ ,
\end{align}
\end{subequations}
with $\varepsilon^{\nu\lambda k_1 k_2}=\varepsilon^{\nu\lambda \rho \sigma}k_{1\rho}k_{2\sigma}$ and $\text{C}_0(m^2)\equiv\text{C}_0(k_1^2,k_2^2,(k1+k_2)^2,m^2,m^2m^2)$ the triangular Passarino-Veltman function. This well-known result exhibits the freedom that one has to dispatch the anomaly between the three currents. For example, if we want the vector currents to be conserved (as when they are to be coupled to photons), then the vector Ward identity $\partial_{\mu}J_V^{\mu}=0$ must hold, which requires to set $a=b=0$.

The third Ward identity is that for the axial current. Actually, we can identify in the first term on the right the triangle amplitude for a pseudoscalar and two vector currents, 
\begin{equation}
    T_{PVV}^{\mu\nu}=-\frac{im}{2\pi^2}\text{C}_0(m^2)\varepsilon^{\mu\nu k_1 k_2},
\label{PVVC0}
\end{equation}
thus,
\begin{equation}
     i(k_{1}+k_{2})_{\lambda}T_{AVV}^{\lambda\mu\nu}=2imT_{PVV}^{\mu\nu}-\frac{(a+b-2)}{4\pi^2}\varepsilon^{\mu\nu k_1 k_2}.
\end{equation}
Therefore, if we want the classical axial ward identity $\partial_{\mu}J_A^{\mu}=2imP$ with $P=\bar{\psi}\gamma^5\psi$ to hold, we have to choose $a+b=2$, that is inconsistent with the choice $a=b=0$ necessary to conserve the vectorial currents.

We want to emphasize a nuance here: it is not because a Ward identity is broken that there is systematically an effective non-zero contribution emerging from the $AVV$ triangle. We see this explicitly when we perform an expansion in $1/m$ and consider only the leading-order terms, affected by the anomaly. Indeed, in the $m\rightarrow \infty$ limit:
\begin{equation}
i(k_{1}+k_{2})_{\lambda}T_{AVV}^{\lambda\mu\nu}\underset{m\rightarrow\infty}{=}-\frac{(a+b)}{4\pi^2}\varepsilon^{\mu\nu k_1 k_2}.
\end{equation}
We see that upon our choice $a=b=0$, that preserves the vector ward identity, there is no mass-independent contribution from the axial divergence: this is the \textit{Sutherland-Veltman} theorem \cite{Sutherland:1967vf,Veltman:1967}. In other words, when the fermion is massive, the $AVV$ triangle amplitude starts at $\mathcal{O}(m^{-2})$. This leaves at $\mathcal{O}(m^0)$ a correspondence between the pseudoscalar triangle amplitude $2 i m T_{PVV}^{\mu\nu}$ and the anomaly. Let us stress three points though. First, the pseudoscalar amplitude is not anomalous, hence this correspondence is purely parametric. Second, the theorem holds only in the $m\rightarrow  \infty$ limit, and breaks down for massless fermions. In that latter case, there is no pseudoscalar triangle yet the anomaly is of course still there. Third, finally, this holds only when the vector currents are conserved, and the anomaly is entirely set on the axial current.

It is also possible to mass-expand the amplitude of the $AVV$ triangle directly (that is the quantity of interest for any 3-form coupling). At the leading order:
\begin{equation}
    T_{AVV}^{\lambda\mu\nu}\underset{m\rightarrow\infty}{=}\frac{-i}{4\pi^2}\left(a\;\varepsilon^{\lambda\mu\nu k_1}-b\;\varepsilon^{\lambda\mu\nu k_2}\right) \ .
\label{AVVgen}
\end{equation}
Following the prescription that ensures the vectorial Ward identities $a=b=0$, the mass-independent contribution of the $AVV$ amplitude vanishes, as expected from the Sutherland-Veltman theorem. Therefore, the preservation of the vectorial Ward identities translates automatically into an impossibility to mediate any process directly through the $AVV$ triangle.

Though the  $a=b=0$ configuration is that most often encountered, since vector currents are usually coupled to photons, the framework of dark higher-forms will prove to be less restrictive. Actually, some processes involving higher-form fields do require to set the anomaly on one of the vector current, even when it is to be coupled to photons or gluons. Such anomalous breaking of the QED or QCD gauge invariance are however very specific, and provide a complementary view on the $AVV$ process. The reason why this makes sense physically has to do with the properties of the $AVV$ amplitude itself, instead of its divergences.

So, let us imagine we are in the $a=b=1$ configuration that preserves the axial Ward identity:
\begin{equation}\label{AVVaxial}
    T_{AVV}^{\lambda\mu\nu}\underset{m\rightarrow\infty}{=}\frac{-i}{4\pi^2}\left(\varepsilon^{\lambda\mu\nu k_1}-\varepsilon^{\lambda\mu\nu k_2}\right)\;\;\;\;\text{for }\;a=b=1 \;.
\end{equation}
The amplitude does not vanish at $\mathcal{O}(m^0)$ since we are not in a configuration where the Sutherland-Veltman theorem holds, and the classical vector Ward identities are broken. Yet, the amplitude thus obtained is very specific: at the level of the corresponding effective operator, we can write
\begin{equation}\label{Gdef}
      T_{AVV}^{\lambda\mu\nu}\;\epsilon^{*}_{\mu}(k_1)\epsilon^{*}_{\nu}(k_2)\rightarrow\frac{i}{8\pi^2}\varepsilon^{\mu\nu\rho\lambda}A_{\mu}F_{\nu\rho}\equiv \frac{i}{3}\varepsilon^{\mu\nu\rho\lambda}G_{\mu\nu\rho}\;,
\end{equation}
where $\epsilon^{*}_{\mu}(k_1)$ and $\epsilon^{*}_{\nu}(k_2)$ are the outgoing polarization vectors and we have set the gauge coupling constant to one. The quantity 
\begin{equation}
G_{\lambda} = \frac{-1}{3!}\varepsilon_{\mu\nu\rho\lambda}G^{\mu\nu\rho}\;,\;\;\partial^{\mu} G_{\mu} =\frac{1}{16\pi^2}F_{\mu\nu}\tilde{F}^{\mu\nu}\;,
\end{equation}
is the \textit{Chern-Simons current}, and $G_{\mu\nu\rho}$ the corresponding dual 3-form. It is here written in terms of abelian gauge fields, but the same result is obtained with gluons (though one would then need to consider also four and five-point diagrams).

Now, a general theorem states that under gauge transformations, the Chern-Simons 3-form receives two terms (see e.g. Ref.~\cite{Baez:1995sj}). The first is a total derivative, and corresponds to so-called small gauge transformations (those homotopic to the identity). In the abelian case, clearly, the small gauge transformation $A_{\mu}\rightarrow A_{\mu}+\partial_{\mu}\chi$ induces
\begin{equation}
\delta G_{\lambda}=\frac{1}{16\pi^2}\varepsilon_{\mu\nu\rho\lambda}\partial^{\mu}{\chi}F^{\nu\rho}\;,
\end{equation}
but this remains true for non-abelian gauge fields. Thanks to this property, and after partial integration (assuming as usual that $F^{\nu\rho}$ vanishes at infinity), the $AVV$ triangle provides a suitable gauge invariant coupling provided the axial current satisfies
\begin{equation}
\varepsilon_{\lambda\mu\nu\rho}\partial^{\mu}J_A^{\lambda} = 0\;.
\label{invcond}
\end{equation}
There are two solutions:
\begin{equation}
J_A^{\lambda} = \partial^{\lambda}\Phi\;\;\;,\;\;J_A^{\lambda} =  \frac{-1}{3!}\varepsilon^{\mu\nu\rho\lambda} C_{\mu\nu\rho} \; .
\label{Ja2poss}
\end{equation}
The first one is the usual current for a pseudoscalar field $\Phi$, as relevant e.g. for the $\pi^0\gamma\gamma$ vertex. Importantly, its axial form emphasizes the shift-symmetric nature of the corresponding $\Phi F_{\mu\nu}\tilde{F}^{\mu\nu}$ coupling, i.e., its invariance if $\Phi\rightarrow \Phi + \Lambda$, as appropriate whenever $\Phi$ is a (pseudo) Goldstone boson. The second possibility involves the 3-form field $C_{\mu\nu\rho}$, with the condition of Eq.~(\ref{invcond}) satisfied thanks to the Lorenz condition $\partial^{\mu} C_{\mu\nu\rho} = 0$. In practice, the $AVV$ triangle then couples directly the 3-form field to the Chern-Simons three form, $C_{\mu\nu\rho}G^{\mu\nu\rho}$. Technically, this is the dual description of the shift-symmetric $\Phi$ coupling to $F_{\mu\nu}\tilde{F}^{\mu\nu}$. 

Ultimately, this dual picture shows that at the level of the $AVV$ amplitude, the anomaly has to be put in the vector current to couple to the Chern-Simons current. This is to be contrasted to the usual interpretation for $J_A^{\lambda} = \partial^{\lambda}\Phi$. In that case, if one integrates by part the fermionic coupling $\partial_{\lambda}\Phi \bar{\psi} \gamma^{\lambda}\gamma^5 \psi = 2 i m \Phi \bar{\psi} \gamma^5 \psi$, one recovers again the $\Phi F_{\mu\nu}\tilde{F}^{\mu\nu}$ coupling via a non-anomalous pseudoscalar loop. Yet, it is not the same to integrate by part before or after the loop integration. Also, this interpretation breaks down for a 3-form field, since there is no way to integrate by part a $\varepsilon_{\mu\nu\rho\lambda}C^{\mu\nu\rho}\bar{\psi} \gamma^{\lambda}\gamma^5 \psi$ coupling. In Section~\ref{phiCGG}, we will compare the phenomenological consequences in more details. 

The physical picture one gets is particularly intuitive in the non-abelian case. The Chern-Simons form for those gauge groups do vary under large gauge transformations, but it does so in a very constrained way, shifting in integer steps according to the winding number of the gauge transformation. For the $AVV$ triangle to sense the underlying topology of the gauge group, the anomaly has to be put on one of the gauge boson leg. In this context, it should be noted that the Chern-Simons 3-form $G^{\mu\nu\rho}$ inherits its dynamics from the QCD topological susceptibility, and essentially act as a background electric field~\cite{Luscher:1978rn,DiVecchia:1980yfw, Aurilia:1980xj}. Again, it is via its vector current leg that the $AVV$ triangle feels this background. Finally, it should be mentioned that this interpretation has been exploited to construct alternative axion models, see e.g.  Refs.~\cite{dvali2005threeformgaugingaxionsymmetries,dvali2022strongcpgravity}.

To close this section and for future use, performing a similar calculation for all the other possible scalar and pseudoscalar triangle amplitudes gives the generic pattern:
\begin{equation}%
\begin{array}
[c]{llllllll}%
T_{SSS}\sim & 1/\varepsilon\ ,\ \ \  & T_{SVV}\sim & \text{finite}\ ,\ \ \  &
T_{PSS}\sim & 0\ ,\ \ \  & T_{PVV}\sim & \text{finite}\ ,\\
T_{SSP}\sim & 0\ , & T_{SVV}\sim & 0\ , & T_{SSP}\sim & 0\ , & T_{PVA}\sim &
0\ ,\\
T_{SPP}\sim & 1/\varepsilon\ , & T_{SVV}\sim & 1/\varepsilon\ , & T_{PPP}\sim
& 0\ , & T_{PAA}\sim & \text{finite}\ ,
\end{array}
\label{SSPPtriangles}%
\end{equation}
with also $T_{APP}=T_{ASS}=T_{ASP}=0$ and $T_{VPP}=T_{VSS}=T_{VSP}=0$. The three finite cases correspond to the usual amplitudes%
\begin{equation}
T_{SVV}^{\mu\nu}=\frac{1}{6\pi^{2}m}(g^{\mu\nu}k_{1}\cdot k_{2}-k_{2}^{\mu
}k_{1}^{\nu})\ ,\ \ T_{PVV}^{\mu\nu}=3T_{PAA}^{\mu\nu}=\frac{1}{4\pi^{2}%
m}e^{\mu\nu\rho\sigma}k_{1,\rho}k_{2,\sigma}\ .
\label{PVVPAA}%
\end{equation}
Notice that $T_{PVV}^{\mu\nu}$ as given above is simply Eq.~(\ref{PVVC0}) expanded to first order in $1/m$. The reasons why many of these amplitudes vanish and the interpretation of those that diverge will be discussed in the next section.

\subsection{Triangles involving $T$ and $\tilde{T}$ couplings}

In the SM, there is no renormalizable tensorial coupling between bosons and fermions, the leading magnetic couplings arising at dimension five. By contrast, the 2-form field couples to the fermions through the renormalizable operators
\begin{subequations}
\begin{align}
    \mathcal{L}_{B T}&=B_{\mu\nu}\bar{\psi}\sigma^{\mu\nu}\psi\;,\\
    \mathcal{L}_{B\tilde{T}}&=B_{\mu\nu}\bar{\psi}\sigma^{\mu\nu}\gamma^5\psi\;.
\end{align}
\label{BTfermion}
\end{subequations}
In a theory involving a 2-form field, it is thus relevant to consider the contributions of triangle diagrams involving $T$ and $\tilde{T}$. For example, a 2-form field could in principle couple to two photons through a $TVV$ or a $\tilde{T}VV$ triangle, or if light enough, pions could decay into two 2-form fields via the $ATT$ diagram.

Yet, we will see in the following that very few of these processes exist because of Furry's theorem~\cite{Furry:1937zz}, i.e.,  the charge-conjugation structure of the underlying triangle amplitudes. In the following, we will investigate the triangles that contain at least one $T$ or $\tilde{T}$ vertex, with other vertices being either $V$ or $A$. These investigations will also lead us to consider certain triangles containing a $P$ vertex, necessary to disentangle and isolate the possible emerging anomalies.

\subsubsection{Forbidden triangles and the Furry theorem}

To set the stage, let us recall why the $VVV$ triangle contribution is exactly null from a charge-conjugation point-of-view. To do so, we remind the $VVV$ triangle amplitude,
\begin{equation}
    T_{VVV}^{\lambda\mu\nu}=i\int\frac{d^4q}{(2\pi)^4}\text{Tr}\left[S(q-k_1)\gamma^{\lambda}S(q)\gamma^{\mu}S(q+k_2)\gamma^{\nu}\right]+\{\mu\leftrightarrow\nu,k_1\leftrightarrow k_2\}\;.
\end{equation}
One can insert the unitary charge conjugation operator $C$ between each factor of the expression (as $C^{-1}C=\mathbb{I}$). Then, using the transformations $C\gamma^{\mu}C^{-1}=-(\gamma^{\mu})^{T}$ and $CS(q)C^{-1}=S(-k)^{T}$,
the properties of the matrix transposition, the cyclicity of the trace and a proper redefinition of the integration measure, one can prove:
\begin{equation}\label{Furry}
    T_{VVV}^{\lambda\mu\nu}=(-1)^3\times T_{VVV}^{\lambda\mu\nu}\Rightarrow T_{VVV}^{\lambda\mu\nu}=0\;.
\end{equation}
If one replaces a vector coupling by an axial coupling, that transforms as $C\gamma^{\mu}\gamma^{5}C^{-1}=+(\gamma^{\mu}\gamma^{5})^{T}$, Furry theorem does not forbid the amplitude anymore, as detailed in the previous section.

We can apply the same reasoning concerning the triangles involving $T$ and $\tilde{T}$ vertices. The key point is to notice that they transform just like a vector coupling under charge conjugation:
\begin{equation}
    C\sigma^{\mu\nu}C^{-1}=-(\sigma^{\mu\nu})^{T}\;\;,\;\;\;\;\;\;\;\;\;\;\;\;\;\;\;\;C\sigma^{\mu\nu}\gamma^5C^{-1}=-(\sigma^{\mu\nu}\gamma^5)^{T}\;.
\end{equation}
Therefore, one could repeat the above demonstration replacing $\gamma^{\mu}$ by $\sigma^{\mu\rho}$ or $\sigma^{\mu\rho}\gamma^5$. It shows that one can apply Furry theorem to any triangle containing an odd number of $V$,$T$ or $\tilde{T}$ couplings. To put things differently, every triangle that does not contain an odd number of $A$ vertices vanishes identically. From that, we conclude that only five amplitudes survive: $ATT$, $A\tilde{T}T$, $A\tilde{T}\tilde{T}$, $AVT$, $AV\tilde{T}$.

\subsubsection{Explicit calculation: UV-divergences and renormalizability}

We now aim at calculating the amplitudes of these five surviving triangles, focusing on the leading order of the $(1/m)$ expansions, and on the possible anomalous nature of these triangles.

We begin with the $AVT$ triangle. To set the stage, let us add to the 2-form couplings to fermions, Eq.~\eqref{BTfermion}, the typical pseudoscalar meson derivative couplings to the fermions:
\begin{equation}
\partial_{\mu}\Phi\bar{\psi}\gamma^{\mu}\psi \;\;,\;\;\; \partial_{\mu}\Phi\bar{\psi}\gamma^{\mu}\gamma^{5}\psi\;. 
\end{equation}
At one loop, one would thus expect $\Phi\rightarrow \gamma B$ to occur. Its amplitude reduces to the $AVT$ triangle:
\begin{equation}
    T_{AVT}^{\lambda,\mu\nu,\alpha}=i\int\frac{d^4q}{(2\pi)^4}\text{Tr}\left[S(q-k_1)\gamma^{\lambda}\gamma^5S(q)\sigma^{\mu\nu}S(q+k_2)\gamma^{\alpha}\right]+\{\mu,\nu\leftrightarrow\alpha,k_1\leftrightarrow k_2\}\;.
\end{equation}
For now, we keep the BMHV scheme to treat the $\gamma^5$ but, unlike the previous subsection, we will choose for simplicity to keep $\gamma^5$ on the first vertex. This prescription enforces the conservation of the vector ward identity, and we check that $k_{2\alpha}T_{AVT}^{\lambda,\mu\nu,\alpha}=0$. Due to the Dirac structure of the trace, the only term of the amplitude that survives is proportional to $m$, that happens to respect:
\begin{equation}
     T_{AVT}^{\lambda,\mu\nu,\alpha}\underset{m\rightarrow\infty}{=}0 \ .
\end{equation}
We see that Sutherland-Veltman theorem holds. Consequently, $(k_1+k_2)_{\lambda}T_{AVT}^{\lambda,\mu\nu,\alpha}\underset{m\rightarrow\infty}{=}0$. As the axial vertex is still following a Ward identity of the form:
\begin{equation}\label{wardAVT}
    i(k_1+k_2)_{\lambda}T_{AVT}^{\lambda,\mu\nu,\alpha}= 2im\;T_{PVT}^{\mu\nu,\alpha}+A_{no}^{\mu\nu,\alpha}\;,
\end{equation}
with $A_{no}^{\lambda,\mu\nu,\alpha}$ the anomalous term, we deduce in the present configuration:
\begin{equation}\label{PVT}
   2im\;T_{PVT}^{\mu\nu,\alpha}\underset{m\rightarrow\infty}{=}-A_{no}^{\mu\nu,\alpha}\;.
\end{equation}
Therefore, we can focus on the $PVT$ triangle in order to identify the anomalous term and the possible underlying effective couplings. We can follow the same prescription to calculate $PVT$, using the BMHV scheme and the Passarino-Veltman reduction in $D=4-2\epsilon$ dimensions. Doing so, the $PVT$ amplitude exhibits a UV-pole of the form:
\begin{equation}
    T_{PVT}^{\mu\nu,\alpha}\subset\frac{1}{4\pi^2\epsilon}\varepsilon^{\mu\nu\alpha k_2}\;.
\label{PVTdiv}
\end{equation}
This is in stark contrast with the usual $AVV$, $SVV$, $PVV$, $AAA$ amplitudes which are all UV-finite. Yet, this feature should have been expected from the start because having $\Phi$, photons, and a 2-form field at hand, nothing prevents the renormalizable coupling
\begin{equation}
    \mathcal{L}_{PVT}=c_{PVT}\Phi F^V_{\mu\nu}\tilde{B}^{\mu\nu}\ ,
\end{equation}
where $F^V_{\mu\nu}$ is the field strength of the vector gauge field $V$. In other words, it is necessary to include this coupling to absorb the divergence in Eq.~(\ref{PVTdiv}). Phenomenologically, this means the strength of the $AVT$ triangle cannot be fixed, but has to be determined experimentally.

Alternatively, one can use the \textit{Pauli-Villard} \cite{Pauli:1949zm} regularization scheme to calculate the amplitude. The procedure has the advantage of allowing the calculation of the Dirac trace and the $\gamma^5$ manipulations at $D=4$. Yet, one should remember that it automatically enforces the vector Ward identities and Sutherland-Veltman theorem. Within this scheme, the UV-pole takes the form of a logarithmic divergence $\text{ln}(m/M)$ with $M$ the mass of the regulator, which has to be absorbed into $c_{AVT}$. Using the Pauli-Villard scheme, it is also trivial to obtain the $PV\tilde{T}$ amplitude from that for $PVT$ since in four dimensions, the relation $\sigma^{\mu\nu}\gamma^{5}=(i/2)\varepsilon^{\mu\nu\rho\sigma}\sigma_{\rho\sigma}$ is valid. Unsurprisingly, we find the same divergence pattern: 
\begin{align}
     T_{AV\tilde{T}}^{\lambda,\mu\nu,\alpha}&\underset{m\rightarrow\infty}{=}0\;,\\
     T_{PV\tilde{T}}^{\mu\nu,\alpha}&\;\;\subset\;\frac{i}{4\pi^2\epsilon}\left(g^{\alpha\nu}k_2^{\mu}-g^{\alpha\mu}k_2^{\nu}\right)\rightarrow \mathcal{L}_{PV\tilde{T}}=\frac{i}{4\pi^2\epsilon}\Phi F^V_{\mu\nu}B^{\mu\nu}\;.
\end{align}
As for $AVT$, the $AV\tilde{T}$ amplitude satisfies the Sutherland-Veltman theorem, its associated pseudoscalar triangle $PV\tilde{T}$ exhibits a UV-pole, and there is a renormalizable operator that needs to be present.

A similar, yet trickier case is the calculation of the $ATT$ amplitude:
\begin{equation}
    T_{ATT}^{\lambda,\mu\nu,\alpha\beta}=i\int\frac{d^4q}{(2\pi)^4}\text{Tr}\left[S(q-k_1)\gamma^{\lambda}\gamma^5S(q)\sigma^{\mu\nu}S(q+k_2)\sigma^{\alpha\beta}\right]+\{\mu,\nu\leftrightarrow\alpha,\beta,k_1\leftrightarrow k_2\} \;.
\end{equation}
Using the usual BMHV scheme, one obtains the UV divergent result:
\begin{align}
T_{ATT}^{\lambda,\mu\nu,\alpha\beta}\subset\frac{i}{12\pi^2\epsilon}\bigg( &- g^{\lambda \nu} \varepsilon^{\alpha \beta \mu k_1} +g^{\lambda \mu} \varepsilon^{\alpha \beta \nu k_1} - g^{\beta \lambda} \varepsilon^{\alpha \mu \nu k_2} + g^{\alpha \lambda} \varepsilon^{\beta \mu \nu k_2}+ (k_1+k_2)^{\lambda} \varepsilon^{\alpha \beta \mu \nu}\nonumber\\& + k_1^{\nu} \varepsilon^{\alpha \beta \lambda \mu} - k_1^{\mu} \varepsilon^{\alpha \beta \lambda \nu} + k_2^{\alpha} \varepsilon^{\beta \lambda \mu \nu} - k_2^{\beta} \varepsilon^{\alpha \lambda \mu \nu} \bigg) \ .
\end{align}
To make sense of this divergence, one has to think in terms of the corresponding effective couplings of an axial current to two outgoing 2-form fields, which respect the Lorenz condition $\partial_{\mu}B^{\mu\nu}=\partial_{\nu}B^{\mu\nu}=0$. The terms of the divergence proportional to $k_1^{\mu}$, $k_1^{\nu}$, $k_2^{\alpha}$ and $k_2^{\beta}$ cannot contribute physically. The rest corresponds to the Feynman rule of the operator:
\begin{equation}\label{ABB}
\mathcal{L}_{ATT}=\frac{c_{ATT}}{2!}J_A^{\lambda}\tilde{B}^{\mu\nu}F^B_{\lambda\mu\nu}\; ,
\end{equation}
which is renormalizable if $J_A^{\lambda}$ is coupled to a dimension-one axial field $A^{\lambda}$. Therefore, like for the $PVT$ case, we can absorb the UV-divergence of the triangle into $c_{ATT}$, which thus has to be included for consistency but whose value is free. Notice also that apart from its UV-pole, the $ATT$ respects Sutherland-Veltman theorem, so only the contribution from $c_{ATT}$ matters.

The axial current is to be coupled to SM particles, and in the present context, this means to pseudoscalar mesons. It thus takes the form $J_A^{\lambda} = \partial^{\lambda}\Phi / \Lambda$, see Eq.~(\ref{Ja2poss}). In that case, integrating by part, $\mathcal{L}_{ATT}$ is equivalent to
\begin{equation}\label{ATT2}
    \mathcal{L}_{ATT}=\frac{c_{ATT}}{2!\Lambda}\partial^{\lambda}\Phi\tilde{B}^{\mu\nu}F^B_{\lambda\mu\nu} = -\frac{c_{ATT}}{3!\Lambda} \Phi F^B_{\mu\nu\rho}\tilde{F}^{B,\mu\nu\rho}\; .
\end{equation}
Whatever the form, these operators induce processes of the form $\Phi\rightarrow BB$. 

A special feature in this case though is that this is not the leading operator, since we can construct 
\begin{equation}\label{PTT2}
    \mathcal{L}_{PTT} = \frac{c_{PTT} \Lambda}{2!} \Phi B_{\mu\nu}\tilde{B}^{\mu\nu}\;,
\end{equation}
which is of dimension three. The $ATT$ triangle does not give any contribution to that operator. Yet, if one first integrates by part the $\Phi$ coupling to fermions, $\partial_{\mu}\Phi\bar\psi \gamma^{\mu}\gamma^5 \psi \rightarrow 2 i m \Phi\bar \psi \gamma^5 \psi$, one encounters the $PTT$ triangle, whose amplitude is
\begin{equation}
    T_{PTT}^{\mu\nu,\alpha\beta}=-i\int\frac{d^4q}{(2\pi)^4}\text{Tr}\left[S(q-k_1)\gamma^5S(q)\sigma^{\mu\nu}S(q+k_2)\sigma^{\alpha\beta}\right]+\{\mu,\nu\leftrightarrow\alpha,\beta,k_1\leftrightarrow k_2\} \; .
\end{equation}
In the BMHV scheme, we again encounter a divergence:
\begin{equation}
    T_{PTT}^{\mu\nu,\alpha\beta}\subset \frac{im}{2\epsilon\pi^2}\varepsilon^{\mu\nu\alpha\beta} \; ,
\end{equation}
which requires the presence of $c_{PTT}$. 

Overall, the situation bears some resemblance to the case of the $AVV$ triangle in the Pauli-Villars scheme, where Sutherland-Veltman holds. Remember that in that case, the $AVV$ triangle vanishes, but one has to include the $PVV$ triangle, which is parametrically accounting for the anomaly in the sense that $\partial^{\mu}A_{\mu}-2im P$ is equal to the anomalous term. Here, the $ATT$ triangle is a pure counterterm, to which one has to add the $PTT$ triangle. For both terms, we need specific counterterms, $c_{ATT}$ and $c_{PTT}$. However, the $c_{PTT}$ is dimensionally enhanced and dominates, exactly like the $PVV$ loop is much larger than the $AVV$ one (since its leading $\mathcal{O}(m^0)$ term vanishes). 

A major difference though is that $PVV$ does not match onto a new dimension-three operator since $\Phi V_{\mu}V^{\mu}$ has the wrong parity, and worse, would break the vector gauge invariance. By contrast, in the tensor case, since we do not impose gauge invariance and started from gauge-breaking couplings to fermions (see Eq.~(\ref{BTfermion})), and thanks to the coincidental possibility in four dimensions to form an antisymmetric contraction of two 2-form fields, the $PTT$ term is not anomalous and can be absorbed into a counterterm. In this respect, the $AVT$ triangle discussed previously is a kind of middle ground. It has to vanish because there is no way to construct a parity-odd coupling since $J_A^{\mu}V^{\nu}\tilde{B}_{\mu\nu}$ would again break the vector gauge invariance, but there is a counterterm for the $PVT$ loop because of the existence of $\Phi F_{\mu\nu}^V\tilde{B}^{\mu\nu}$. All these situations are summarized in Table~\ref{TableT}.

In practice, since all the counterterms have to be determined experimentally anyway, their dynamical generation via triangle graphs is rather irrelevant. Yet, we find it quite instructive from a theoretical point of view to see how the anomalous axial Ward identity is realized in the tensor cases.

\renewcommand{\arraystretch}{1.6}
\begin{table}
    \centering
    \begin{tabular}{|c|c|c|c|c|c|}
    \hline
        Triangle  & UV-divergence & Counterterm & Related to & UV-divergence & Counterterm \\
    \hline
         $AVV$& No & $\varnothing$ & $PVV$ & No & $\varnothing$\\
         \hline
         $AVT$& No & $\varnothing$ & $PVT$ & Yes & $\Phi F^V_{\mu\nu}\tilde{B}_{\mu\nu}$\\
        $AV\tilde{T}$& No & $\varnothing$ & $PV\tilde{T}$ & Yes & $\Phi F^V_{\mu\nu}B_{\mu\nu}$\\    \hline
        $ATT/A\tilde{T}\tilde{T}$& Yes & $A^{\lambda}\tilde{B}^{\mu\nu}F^B_{\lambda\mu\nu}$ & $PTT/P\tilde{T}\tilde{T}$ & Yes & $\Phi B_{\mu\nu}\tilde{B}_{\mu\nu}$ \\
        $AT\tilde{T}$& Yes & $A^{\lambda}B^{\mu\nu}F^B_{\lambda\mu\nu}$  & $PT\tilde{T}$ & Yes & $\Phi B_{\mu\nu}B_{\mu\nu}$ \\
    \hline
    \end{tabular}
    \caption{Dimension-three and four counterterms for the various triangle amplitudes, where $F^V_{\mu\nu}$ is the field strength of the vector gauge field $V$, and $A^{\lambda}$ stands for the axial current, with the phenomenological identification $A^{\lambda} \rightarrow \partial^{\lambda}\Phi/\Lambda$. Thanks to Sutherland-Veltman, the axial amplitudes actually vanish in the first three lines. The only true anomaly is in the difference between the divergence of $AVV$ and $2im$ times $PVV$. Phenomenologically, the pseudoscalar loop contributions always dominates, for dimensional reasons in the last four lines, and because it is parametrically equal to the anomaly in the first line.}
    \label{TableT}
\end{table}
\normalsize

To conclude this section, let us directly write down the results for the $AT\tilde{T}$ and $A\tilde{T}\tilde{T}$ cases. Working with the Pauli-Villard regularization scheme, so that the four-dimensional identity $2\sigma^{\mu\nu}\gamma^{5}=i\varepsilon^{\mu\nu\rho\sigma}\sigma_{\rho\sigma}$ holds, the $A\tilde{T}\tilde{T}$ case is identical to $ATT$ whereas $AT\tilde{T}$ is related to $ATT$ through Hodge-duality:
\begin{subequations}
\begin{align}
    T_{AT\tilde{T}}^{\lambda,\mu\nu,\alpha\beta}\subset &\; \frac{1}{12\pi^2\epsilon}\bigg(k_1^{\alpha}\big(g^{\beta\mu} g^{\lambda\nu}-g^{\beta\nu} g^{\lambda\mu}\big) + k_1^{\beta}\big(g^{\alpha\nu} g^{\lambda\mu}-g^{\alpha\mu} g^{\lambda\nu}\big)+k_2^{\mu}\big(g^{\alpha\lambda} g^{\beta\nu}-g^{\alpha\nu} g^{\beta\lambda}\big)\nonumber\\&\;\;\;\;\;\;\;\;\;\;\;\;\;+ k_2^{\nu}\big(g^{\alpha\mu} g^{\beta\lambda}-g^{\alpha\lambda} g^{\beta\mu}\big)+(k_1^{\lambda}-k_2^{\lambda})\big(g^{\alpha\mu} g^{\beta\nu}-g^{\alpha\nu} g^{\beta\mu}\big)\bigg)\nonumber\\  \rightarrow  &\; \mathcal{L}_{AT\tilde{T}}=\frac{1}{6\pi^2\epsilon}A^{\lambda}B^{\mu\nu}F^B_{\lambda\mu\nu} \ , \\
    T_{A\tilde{T}\tilde{T}}^{\lambda,\mu\nu,\alpha\beta}\subset & \; \frac{i}{12\pi^2\epsilon}\bigg(g^{\lambda \nu} \varepsilon^{\alpha \beta \mu k_1} +g^{\lambda \mu} \varepsilon^{\alpha \beta \nu k_1} - g^{\beta \lambda} \varepsilon^{\alpha \mu \nu k_2} + g^{\alpha \lambda} \varepsilon^{\beta \mu \nu k_2}+ (k_1+k_2)^{\lambda} \varepsilon^{\alpha \beta \mu \nu}\bigg)\nonumber\\\rightarrow  &\;  \mathcal{L}_{A\tilde{T}\tilde{T}}=\frac{i}{6\pi^2\epsilon}A^{\lambda}\tilde{B}^{\mu\nu}F^B_{\lambda\mu\nu} \ .
\end{align}
\end{subequations}
Both $AT\tilde{T}$ and $A\tilde{T}\tilde{T}$ respect Sutherland-Veltman theorem after renormalization. To this, we then have to add the following pseudoscalar triangle amplitudes:
\begin{subequations}
\begin{align}
     T_{PT\tilde{T}}^{\mu\nu,\alpha\beta}&\subset \frac{m}{2\epsilon\pi^2}\left(g^{\alpha\mu}g^{\beta\nu}-g^{\alpha\nu}g^{\beta\mu}\right)\rightarrow    \mathcal{L}_{PT\tilde{T}}=\frac{-im}{2\epsilon\pi^2}\Phi B_{\mu\nu}B^{\mu\nu}\ ,\\
     T_{P\tilde{T}\tilde{T}}^{\mu\nu,\alpha\beta}&\subset \frac{im}{2\epsilon\pi^2}\varepsilon^{\mu\nu\alpha\beta}\rightarrow    \mathcal{L}_{P\tilde{T}\tilde{T}}=\frac{m}{2\epsilon\pi^2}\Phi B_{\mu\nu}\tilde{B}^{\mu\nu} \ .
\end{align}
\end{subequations}
As for the $ATT$ and $PTT$ case discussed previously, the pseudoscalar counterterms always dominates phenomenologically.

\section{Phenomenological comparisons}

It has been known for a long time that there exists a deep connection between the massive scalar field and 3-form as well as between the massive vector and 2-form. As long as the fields remain free, the two representations of a spin-zero (spin-one) particle are formally equivalent: the two representations are said to be \textit{dual} to each other, and related as%
\begin{subequations}
\label{DualEqn}%
\begin{align}
C_{\mu\nu\rho}\overset{}{=}-\frac{1}{1!}\frac{1}{m}\varepsilon_{\mu\nu
\rho\sigma}F^{\phi,\sigma}\ \   & \leftrightarrow\ \ \phi=\frac{1}{4!}\frac
{1}{m}\varepsilon^{\mu\nu\rho\sigma}F_{\mu\nu\rho\sigma}^{C}\ ,\\
B_{\mu\nu}\overset{}{=}-\frac{1}{2!}\frac{1}{m}\varepsilon_{\mu\nu\rho\sigma
}F^{A,\rho\sigma}\ \   & \leftrightarrow\ \ A_{\mu}=\frac{1}{3!}\frac{1}%
{m}\varepsilon_{\mu\nu\rho\sigma}F^{B,\nu\rho\sigma}\ , \label{BtoA}
\end{align}
\end{subequations}
with $m$ the mass of the dark field and the field strengths $F_{\mu}^{\phi}=\partial_{\mu}\phi/0!$, $F_{\mu\nu}^{A}=\partial_{\lbrack\mu}A_{\nu]}/1!$, $F_{\mu\nu\rho}^{B}=\partial_{\lbrack\mu}B_{\nu\rho]}/2!$, and $F_{\mu\nu\rho\sigma}^{C}=\partial_{\lbrack\mu}C_{\nu\rho\sigma]}/3!$, see Ref.~\cite{Plantier:2025hcm} for our definitions and conventions. These are massive versions of the usual electromagnetic duality $F_{\mu\nu}=\varepsilon_{\mu\nu\rho\sigma}\tilde{F}^{\rho\sigma}/2!$. When coupled to SM fields however, these dualities are jeopardized. Going from the scalar (vector) basis to the 3-form (2-form) basis of effective interactions involves a shuffling of the orders of mass, with dominant operator in a basis becoming subdominant in the other and vice-versa. Also, duality relations involve the parity-odd epsilon tensor, and thus relate operators of opposite parities. 

In Ref.\cite{Plantier:2025hcm}, we analyzed systematically the effective operator bases for each pair of embeddings, including gauge (or shift) symmetry properties and scale suppression. Here, our goal is to concentrate specifically on cubic bosonic interactions to which fermionic loops may contribute via triangle diagrams. Specifically, we will investigate the phenomenological differences between the decay of a pseudoscalar meson in two dark forms $\Phi_{SM}\rightarrow XX$, with $\Phi_{SM}$ either a Higgs boson, a scalar or a pseudoscalar meson, and the decay of a dark form in two photons $X\rightarrow\gamma\gamma$, with $X=\phi/C$ or $X=A/B$.

Concerning the fermionic triangle loops, we will consider only the renormalizable interactions between higher-form fields and fermions (with and without $\gamma^{5}$) \cite{Plantier:2025hcm}. For clarity, we set all the coupling constants $g_{X\psi}$ to $1$:
\begin{subequations}
\label{couplages_fermions}%
\begin{align}
\mathcal{L}_{\phi\psi} &  =\phi\bar{\psi}\psi\ ,\ \ \phi\bar{\psi}\gamma
^{5}\psi\ ,\\
\mathcal{L}_{A\psi} &  =A_{\mu}\bar{\psi}\gamma^{\mu}\psi\ ,\ \ A_{\mu}%
\bar{\psi}\gamma^{\mu}\gamma^{5}\psi\ ,\\
\mathcal{L}_{B\psi} &  =B_{\mu\nu}\bar{\psi}\sigma^{\mu\nu}\psi\ ,\ \ B_{\mu
\nu}\bar{\psi}\sigma^{\mu\nu}\gamma^{5}\psi \label{Bpsipsi} \ ,\\
\mathcal{L}_{C\psi} &  =C_{\mu\nu\rho}\varepsilon^{\mu\nu\rho\sigma}\bar{\psi
}\gamma_{\sigma}\psi\ ,\ \ C_{\mu\nu\rho}\varepsilon^{\mu\nu\rho\sigma}%
\bar{\psi}\gamma_{\sigma}\gamma^{5}\psi \ .
\end{align}
\end{subequations}
Notice that these operators are not invariant under the gauge (or shift) transformation of the dark field. The only exception is $A_{\mu}\bar{\psi}\gamma^{\mu}\psi$, which may arise from a gauge-invariant kinetic mixing of the dark field with the photon field strength (or $U(1)_{Y}$ if constructed at the level of the SM).

\subsection{The dark spin-zero pair productions : $\Phi_{SM}\rightarrow
\phi\phi/CC$}

Because the two spin-zero particles $\phi\phi$ or $CC$ are produced with zero total angular momentum, they are necessarily in a parity even s-wave state. Thus, this process can occur when $\Phi_{SM}$ is a scalar particle, but is forbidden for a pseudoscalar particle. The best illustration of this selection rule is in the neutral kaon system, with the $J^{PC}=0^{++}$ and $0^{-+}$ state identified as $K_{S}$ and $K_{L}$ in the CP-conserving limit, and the dominant decays being $K_{S}\rightarrow\pi\pi$ and $K_{L}\rightarrow\pi\pi\pi$.

\subsubsection{Scalar $\Phi_{SM}$ case}

For a scalar $\Phi_{SM}$, the two processes $\Phi_{SM}\rightarrow\phi\phi$ and $\Phi_{SM}\rightarrow CC$ are totally similar. The dominant operator is of dimension three in both cases,
\begin{equation}
\Lambda\Phi_{SM}\phi\phi\leftrightarrow\Lambda\Phi_{SM}C_{\mu\nu\rho}C^{\mu
\nu\rho}\ ,\label{Scalar1}%
\end{equation}
and dominates over loop contributions from $T_{SSS}/T_{SPP}$ or $T_{SVV}/T_{SAA}$ fermionic triangles. Importantly, all these triangle amplitudes but that of $T_{SVV}$ are divergent, see Eq.~(\ref{SSPPtriangles}), and thus do require the presence of these renormalizable couplings. Notice that if $\Phi_{SM}$ is to stand for the SM Higgs boson, then one should rather start from higher-dimensional operators involving e.g. $H^{\dagger}H\rightarrow(v+h^{0})^{2}$.

If $\Phi_{SM}$ has to be (manifestly, see below) derivatively coupled like the $K_{S}$ would be, then the leading operators are%
\begin{equation}
\frac{1}{\Lambda}\partial_{\mu}\Phi_{SM}\phi\partial^{\mu}\phi\leftrightarrow
\frac{1}{\Lambda}\partial_{\sigma}\Phi_{SM}C_{\mu\nu\rho}\partial^{\sigma
}C^{\mu\nu\rho}\ .\label{Scalar2}%
\end{equation}
No other operator arise upon enforcing the Lorenz condition $\partial^{\mu}C_{\mu\nu\rho}=0$. Concerning fermion loops, they do contribute to these operators, but this is rather irrelevant since the gauge/shift symmetric couplings of $\phi$ and $C$ to fermions are already of dimension five, and one has in any case to include all gauge/shift symmetric operators at each order.

Finally, it should be noted that imposing the shift symmetry for the SM scalar is not equivalent to imposing the invariance under the gauge (or shift) transformations of the dark field. The converse is true though at leading order, because this latter constraint leads to%
\begin{equation}
\frac{1}{\Lambda}\Phi_{SM}\partial_{\mu}\phi\partial^{\mu}\phi\leftrightarrow
\frac{1}{\Lambda}\Phi_{SM}F_{\mu\nu\rho\sigma}^{C}F^{C,\mu\nu\rho\sigma
}\ ,\label{Scalar3}%
\end{equation}
which differs from Eq.~(\ref{Scalar2}) by operators proportional to those in Eq.~(\ref{Scalar1}), scaled by $m^{2}/\Lambda$. Now, any shift $\Phi_{SM}\rightarrow\Phi_{SM}+\Lambda$ with $\Lambda$ constant can be absorbed into a wave-function correction for the dark field, and would thus be unobservable.

In conclusion, when $\Phi_{SM}$ is a scalar, there is no reason to expect any difference in the scaling of the $\Phi_{SM}\rightarrow\phi\phi$ vs. $\Phi_{SM}\rightarrow CC$ decay rates, no matter the underlying assumptions. Let us stress though that if instead of a decay process, one considers a transition like $\Phi_{1,SM}\rightarrow\Phi_{2,SM}\phi\phi$ or $\Phi_{1,SM}\rightarrow\Phi_{2,SM}CC$, then the differential rate are very different, with that for $CC$ peaked at larger invariant mass compared to that for $\phi\phi$, see Ref.~\cite{Plantier:2025hcm}.

\subsubsection{Pseudoscalar $\Phi_{SM}$ case}

For a pseudoscalar, as explained at the beginning of this section, the process is forbidden. Without surprise, it is not possible to construct a cubic coupling involving the $\varepsilon$ tensor, $\Phi_{SM}$ and two identical $\phi$ fields. This would necessarily require at least four derivatives, and two of them would have to act on the same field and vanish by the antisymmetric contraction.

For the $C$ embedding, surprisingly, operators do exist, but they all disappear once imposing the Lorenz condition $\partial^{\mu}C_{\mu\nu\rho}=0$. For example, the lowest-dimensional such operator is of dimension five but by partial integration, we can write%
\begin{equation}
\frac{1}{\Lambda}\partial^{\mu}\Phi_{SM}\partial^{\nu}C_{\alpha\beta\gamma
}\varepsilon^{\alpha\beta\gamma\rho}C_{\mu\nu\rho}\sim\frac{1}{\Lambda}%
\Phi_{SM}\partial^{\nu}C_{\alpha\beta\gamma}\varepsilon^{\alpha\beta\gamma
\rho}\partial^{\mu}C_{\mu\nu\rho}\rightarrow0\ .\label{Scalar4}%
\end{equation}
The cancellation is thus less automatic in the 3-form case.

A perfectly analogous pattern emerges at the loop level. In the $\phi$ case, the triangle amplitudes involving a pseudoscalar coupling $\Phi_{SM}\bar{\psi}\gamma^{5}\Phi$ or an effective $\partial_{\mu}\Phi_{SM}\bar{\psi}\gamma^{\mu}\gamma^{5}\psi$ coupling all vanish thanks to the structure of the Dirac traces, which forces:
\begin{equation}
T_{PSS}=T_{PPP}=T_{ASS}=T_{APP}=0\ .\label{Scalar5}%
\end{equation}
Therefore, no $\Phi_{SM}\phi^{2}$ coupling could emerge from a fermion loop, no matter whether $\Phi_{SM}$ is a pseudoscalar or a pseudo-Goldstone boson.

By contrast, for the 3-form field, it seems possible at the first glance to obtain a non-zero fermion loop contribution to the decay. The underlying triangle amplitudes are that for the $PVV$ or $PAA$ configuration, which are both finite. Focusing on the vector interactions:
\begin{equation}
\mathcal{M}_{\Phi\rightarrow CC}=\;T_{PVV}^{\sigma\eta}\left(  \varepsilon
_{\mu\nu\rho\sigma}\;\epsilon_{(\lambda_{1})}^{\ast\;\mu\nu\rho}%
(k_{1})\right)  \left(  \varepsilon_{\alpha\beta\gamma\eta}\;\epsilon
_{(\lambda_{2})}^{\ast\;\alpha\beta\gamma}(k_{2})\right)  \ ,\label{Scalar6}%
\end{equation}
with $\epsilon_{(\lambda)}^{\ast\;\mu\nu\rho}(k)$ the polarization tensor of the outgoing 3-form field with impulsion $k$ and polarization state $\lambda$. Using the expression of $PVV$ exposed in Eq.~(\ref{PVVPAA}), we find at the first order:
\begin{equation}
\mathcal{M}_{\Phi\rightarrow CC}=\left[  \varepsilon_{\mu\nu\rho\sigma
}\;\varepsilon_{\alpha\beta\gamma\eta}\left(  \frac{i}{4\pi^{2}m}%
\varepsilon^{\sigma\eta k_{1}k_{2}}\right)  \right]  \epsilon_{(\lambda_{1}%
)}^{\ast\;\mu\nu\rho}(k_{1})\;\epsilon_{(\lambda_{2})}^{\ast\;\alpha
\beta\gamma}(k_{2})\ ,\label{Scalar7}%
\end{equation}
which contributes to the effective operator in Eq.~(\ref{Scalar4}), hence vanishes only thanks to the Lorenz condition $\partial^{\mu}C_{\mu\nu\rho}=0$. The reasoning is obviously the same using the axial couplings of the $C$ in view of Eqs.~(\ref{SSPPtriangles}) and~(\ref{PVVPAA}). The situation is also identical in the presence of the effective $\partial_{\mu}\Phi_{SM}\bar{\psi}\gamma^{\mu}\gamma^{5}\psi$ coupling, by virtue of the anomalous Ward identity which forces the $AVV$ and $AAA$ triangles to also match onto the operator in Eq.~(\ref{Scalar4}). Finally, notice that if the $C$ gauge invariance is imposed, the leading couplings to fermions are $\varepsilon^{\mu\nu\rho\sigma}F_{\mu\nu\rho\sigma}^{C}\bar{\psi}\psi$ and $\varepsilon^{\mu\nu\rho\sigma}F_{\mu\nu\rho\sigma}^{C}\bar{\psi}\gamma^{5}\psi$, and both
would automatically lead to vanishing loop contributions according to Eq.~(\ref{Scalar5}), without the need to call in the Lorenz condition.

\subsection{The dark spin-zero decays into photons: $\phi/C\rightarrow\gamma\gamma$}
\label{phiCGG}

When the dark scalar is embedded into a scalar field $\phi$, the leading operators are simply the usual%
\begin{equation}
\mathcal{L}_{\phi\gamma\gamma}=\frac{1}{\Lambda}\phi\mathcal{F}_{\mu\nu
}\mathcal{F}^{\mu\nu}\ ,\ \frac{1}{2\Lambda}\phi\mathcal{F}_{\mu\nu
}\mathcal{\tilde{F}}^{\mu\nu} \ ,\label{phiFF}
\end{equation}
where $\mathcal{F}_{\mu\nu}$ stands for the photon field strength. The corresponding decay rates are:
\begin{equation}
\Gamma(\phi \rightarrow \gamma\gamma) =\frac{m_{\phi}^3}{16\pi\Lambda^2}\ .
\end{equation}
The fermion loops do contribute to them, either via the $SVV$ or $PVV$ triangles, see Eq.~(\ref{PVVPAA}). Further, if the $\phi$ shift symmetry is imposed, then the first operator is forbidden, but the second can arise because it is secretly shift-symmetric, as explained in the first section, see Eq.~(\ref{Ja2poss}). In practice, if $\phi$ is derivatively coupled, then $\phi\rightarrow\gamma\gamma$ involves the $AVV$ triangle which, as explained in the first section, satisfies the Sutherland-Veltman theorem. It thus does not directly contribute to the $\phi\mathcal{F}_{\mu\nu}\mathcal{\tilde{F}}^{\mu\nu}$ operator, but its anomalous piece does, in agreement with the anomalous axial Ward identity. At the end of the day, starting with a pseudoscalar or an axial coupling is totally equivalent.

If the dark scalar is represented by a 3-form field, the situation is drastically different. To start with, there is now a dimension-four operator able to induce directly $C\rightarrow\gamma\gamma$:
\begin{equation}
\mathcal{L}_{C\gamma\gamma}=\lambda C_{\mu\nu\rho}G^{\mu\nu\rho}\ ,
\end{equation}
with $\lambda$ some dimensionless constant and $G^{\mu\nu\rho}$ the QED Chern-Simons form. This is a very peculiar coupling that deserves several comments:

\begin{enumerate}
\item The $C_{\mu\nu\rho}G^{\mu\nu\rho}$ operator is actually dual to $\phi\mathcal{F}_{\mu\nu}\mathcal{\tilde{F}}^{\mu\nu}$, as can be seen from Eq.~(\ref{DualEqn}). Yet, the fact that the duality relations involve a mass scale has serious consequences in an effective framework. Of course, one may actually call in the duality to say that $\lambda\sim m/\Lambda$ with $m$ the dark particle mass, but this is not compulsory. In some sense, $C_{\mu\nu\rho}G^{\mu\nu\rho}$ is the equivalent of the usual kinetic mixing for a dark vector field, for which the coupling also needs to be tiny. Notice though that $C_{\mu\nu\rho}G^{\mu\nu\rho}$ does break the $C$ gauge symmetry.

\item If one removes $C_{\mu\nu\rho}G^{\mu\nu\rho}$, then it comes back at the loop level via the $AVV$ triangle, without any mass suppression. The amplitude is given by:
\begin{equation}
\mathcal{M}_{C\rightarrow\gamma\gamma}=(Qe)^{2}\times\epsilon_{\alpha\beta\gamma
}^{(\lambda_{C})}(k_{1}+k_{2})\left(  \varepsilon_{\alpha\beta\gamma\sigma
}T_{AVV}^{\sigma\mu\nu}\right)  \;\epsilon_{\mu}^{\ast}(k_{1})\epsilon_{\nu
}^{\ast}(k_{2})\ .
\end{equation}
As explained in \textbf{Section 1.1}, if we do not impose the Sutherland-Veltman configuration, then the $\mathcal{O}(m^{0})$ term of $T_{AVV}^{\sigma\mu\nu}$ does not vanish, see Eq.~(\ref{AVVaxial}). Plugged in the above equation, $\mathcal{M}_{C\rightarrow\gamma\gamma}$ matches on the $C_{\mu\nu\rho}G^{\mu\nu\rho}$ operator with%
\begin{equation}
\lambda_{loop}=-\frac{3i(Qe)^{2}}{4\pi^{2}}\ .
\end{equation}
This is actually the only loop-induced effect since the $VVV$ triangle vanishes by Furry's theorem.

\item The subleading terms in $T_{AVV}^{\sigma\mu\nu}$ match onto dimension-six operators, for example into 
\begin{equation}
\varepsilon^{\mu\nu\rho\sigma}F_{\mu\rho\sigma\sigma}^{C}\mathcal{F}_{\alpha\beta}\mathcal{\tilde{F}}^{\alpha\beta}\ ,
\end{equation}
which can be rewitten entirely in terms of $F^{C}$ and $\mathcal{F}$. Those operators become leading if the $C$ gauge invariance is imposed. It must be noted though that no operators truly involving $\mathcal{F\tilde{F}}$ ever arise since the $\varepsilon$ tensor of the loop contracts with the $\varepsilon$ tensor of the $C$ coupling to fermions. Fundamentally, the fermion couplings $\phi\bar{\psi}\gamma^{5}\psi$ and $C_{\mu\nu\rho}\varepsilon^{\mu\nu\rho\sigma}\bar{\psi}\gamma_{\sigma}\gamma^{5}\psi$ have opposite parity properties if one na\"{\i}vely counts the number of $\gamma^{5}$ and the number of $\varepsilon$ tensors. In this sense, the parity property of the $C\gamma\gamma$ coupling is thus inverted compared to that of $\phi\gamma\gamma$.

\item QED gauge invariance is maintained, even though $T_{AVV}^{\sigma\mu\nu}$ as given in Eq.~(\ref{AVVaxial}) naively breaks the usual Ward identity, $k_{1,\mu}T_{AVV}^{\sigma\mu\nu}\neq0$ and $k_{2,\nu}T_{AVV}^{\sigma\mu\nu}\neq0$. To check this, one can compute the decay rate setting the photon polarization sum to%
\begin{equation}
\sum_{\lambda}\epsilon_{\mu}^{\lambda}(k_{1})\epsilon_{\nu}^{\lambda,\ast
}(k_{1})=-g^{\mu\nu}+x\frac{k^{\mu}k^{\nu}}{m^{2}}\ ,
\end{equation}
for arbitrary $x$. One finds then that this parameter is dutifully projected out when summing over the $C$ field polarization states (whose lengthy expression can be found in Ref.~\cite{Plantier:2025hcm}), and one finds the rate:
\begin{equation}
\Gamma(C \rightarrow \gamma\gamma) =\frac{9\; m_C}{256\pi^5} \ .
\end{equation}
This peculiar situation arises because gauge invariance is hidden in the Lorenz condition for the $C$ field, see Eq.~(\ref{Ja2poss}). We are not aware of any other case in which QED gauge invariance is ensured in such an indirect way.

\item All this remains true if gluons replace photons. Even if initially only coupled to quarks, the $C$ field ends up coupled to the QCD Chern-Simons form. In that context, it is tempting to identify the $C$ field to the axion, and this is the step taken e.g. in Ref.~\cite{dvali2005threeformgaugingaxionsymmetries,dvali2022strongcpgravity}. It should be said though that there, the $C$ field gauge invariance is imposed and it is made massive via a Stueckelberg mechanism. At the end of the day, the axion is rather embedded in a 2-form field, because a massless 3-form field propagates no degrees of freedom.
\end{enumerate}

In conclusion, the $\phi$ and $C$ embedding of a dark spin zero particle lead to very different predictions for the two-photon decay, and one would \textit{a priori} expect $\Gamma (C\rightarrow\gamma\gamma)$ to be much larger than $\Gamma(\phi\rightarrow\gamma\gamma)$ since it is induced by a renormalizable operator. On the other hand, if one imposes a dark gauge invariance on the couplings to SM particles, while still allowing for a mass term for the dark state, then $\phi\rightarrow\gamma\gamma$ as induced by the shift-symmetric dimension-five couplings would presumably be much larger than $C\rightarrow\gamma\gamma$ as induced from dimension-six operators.

\begin{figure}
\begin{tikzpicture}[scale=2]
 
  \begin{feynman}
    \vertex (a);
    \vertex [right=1.8 cm of a,label=$\varepsilon_{\alpha\beta\gamma\lambda}\gamma^{\lambda}\gamma^5$](aa);
    \vertex [above=0.05cm of a] (a2);
    \vertex [below=0.05cm of a] (a3);
    \vertex [right=2.5cm of a] (b);
     \vertex [above=0.05cm of b] (b2);
    \vertex [below=0.05cm of b] (b3);
    \vertex [ below right=2.5cm of b] (c);
    \vertex [ below=0.7cm of c,label=$\gamma^{\nu}$] (cc);
    \vertex [ above right=2.5cm of b] (d);
    \vertex [left=0cm of d,label=$\gamma^{\mu}$](dd);
    \vertex [right=2.5cm of c] (e);
    \vertex [right=2.5cm of d] (f);

    \diagram* {
      (a2) -- [ edge label'=\(\epsilon^{\alpha\beta\gamma}(k_1+k_2)\)] (b2),
      (a)--(b);
      (a3)--(b3);
      (b) -- [fermion] (c),
      (c) -- [fermion] (d),
      (d) -- [fermion] (b),
      (d) -- [photon, edge label'=\(\epsilon^{*\mu}(k_1)\)] (f),
      (c) -- [photon,edge label'=\(\epsilon^{*\nu}(k_2)\)] (e),

    };
  \end{feynman}
\end{tikzpicture}
\begin{tikzpicture}[scale=2]
 
  \begin{feynman}
    \vertex (a);
      \vertex [right=1.8 cm of a,label=$\varepsilon_{\alpha\beta\gamma\lambda}\gamma^{\lambda}\gamma^5$](aa);
     \vertex [above=0.05cm of a] (a2);
    \vertex [below=0.05cm of a] (a3);
    \vertex [right=2.5cm of a] (b);
     \vertex [above=0.05cm of b] (b2);
    \vertex [below=0.05cm of b] (b3);
    \vertex [ below right=2.5cm of b] (c);
    \vertex [ below=0.7cm of c,label=$\gamma^{\mu}$] (cc);
    \vertex [ above right=2.5cm of b] (d);
    \vertex [left=0cm of d,label=$\gamma^{\nu}$](dd);
    \vertex [right=3cm of c] (e);
    \vertex [right=3cm of d] (f);
    \vertex [left=1.2cm of e, label=\(\epsilon^{*\nu}(k_2)\)] (ee);
    \vertex [below= 0.5cm, left=1cm of f, label=\(\epsilon^{*\mu}(k_1)\)] (ff);

    \diagram* {
      (a2) -- [ edge label'=\(\epsilon^{\alpha\beta\gamma}(k_1+k_2)\)] (b2),
      (a)--(b);
      (a3)--(b3);
      (b) -- [fermion] (c),
      (c) -- [fermion] (d),
      (d) -- [fermion] (b),
      (d) -- [photon] (e),
      (c) -- [photon] (f),

    };
  \end{feynman}
\end{tikzpicture}
\caption{Triangle graphs inducing $C\rightarrow\gamma\gamma$}
\end{figure}
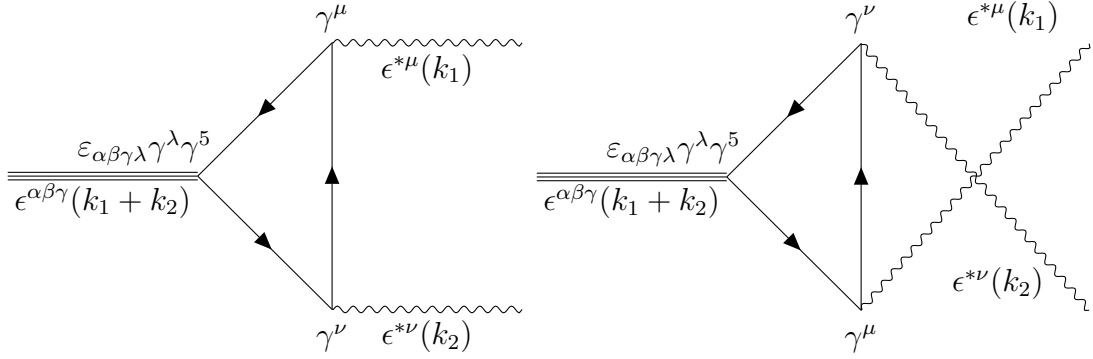
\subsection{The dark spin-one decays into photons: $A/B\rightarrow\gamma\gamma$}

The Landau-Yang theorem~\cite{Landau:1948kw, Yang:1950rg} states unambiguously that no spin-one particle can decay into two on-shell photons. This is a simple consequence of Bose symmetry, which imposes the two photons to be in an even total angular momentum state. If it has been known and demonstrated in the vector case for a long time, it has yet to be done for a 2-form.

The Landau-Yang theorem is a kinematic statement: effective operators susceptible to mediate $A\rightarrow \gamma\gamma$ do exist as no symmetry prevents them in the Lagrangian. It is only at the condition that both the photons and the dark vector are on-shell that these operator contributions vanish. Because of QED invariance, our first operators are of dimension six:
\begin{align}
\mathcal{L}^{\text{dim }6}_{A\gamma\gamma}=&\frac{1}{\Lambda^2}(\partial^{\rho}A_{\rho})\mathcal{F}_{\mu\nu}\mathcal{F}^{\mu\nu}\ ,\ \frac{1}{\Lambda^2}(\partial^{\rho}A_{\rho
})\mathcal{F}_{\mu\nu}\mathcal{\tilde{F}}^{\mu\nu}\ ,\ \frac{1}{\Lambda^2}A_{\rho}\mathcal{F}^{\mu\rho}\partial^{\nu}\mathcal{F}_{\nu\mu}\ ,\ \frac{1}{\Lambda^2}A_{\rho}\mathcal{\tilde{F}}^{\mu\rho}\partial^{\nu}\mathcal{F}_{\nu\mu}\nonumber \ ,\\
&\ \frac{1}{\Lambda^2}A_{\rho}\mathcal{F}^{\mu\nu}\partial_{\mu}F^{\rho}_{\;\nu}\ ,\ \frac{1}{\Lambda^2}A_{\rho}\mathcal{F}^{\mu\nu}\partial_{\mu}\tilde{F}^{\rho}_{\;\nu}\ ,\ \frac{1}{\Lambda^2}A_{\rho}\mathcal{\tilde{F}}^{\mu\nu}\partial_{\mu}F^{\rho}_{\;\nu}\ ,\ \frac{1}{\Lambda^2}\varepsilon_{\mu\nu\rho\sigma}A^{\nu}\mathcal{{F}}^{\rho\alpha}\partial^{\mu}F^{\sigma}_{\;\alpha} \ .
\label{AGG3}
\end{align}
Notice the absence of dark gauge-invariant terms as $F_{\nu}^{\mu}\mathcal{F}_{\mu\rho}\mathcal{F}^{\rho\nu}$ and $F_{\nu}^{\mu}\mathcal{F}_{\mu\rho}\mathcal{\tilde{F}}^{\rho\nu}$ vanish by symmetry. Now, it is easy to see that the first two operators disappear upon the dark vector Lorenz condition whereas the third and fourth ones cancel if one of the photon is on-shell. The last four operators do not trivially cancel, but $|\mathcal{M}(A\rightarrow\gamma\gamma)|^2$ does so for on-shell external states, after summing over the photon and dark vector polarizations. Actually, the tensor structure of $\mathcal{M}(A\rightarrow\gamma\gamma)$ is very constrained by the photon Bose and gauge symmetries~\cite{Zhemchugov_2014}. Labelling momenta as $A^{a}(k_{1}+k_{2})\rightarrow\gamma^{\mu}(k_{1})\gamma^{\nu}(k_{2})$, only four form-factors ever occur:
\begin{align}
\mathcal{M}(A\overset{}{\rightarrow}\gamma\gamma)   = & f_{1}((\varepsilon^{\mu\alpha
\rho\sigma}k_{1}^{\nu}-\varepsilon^{\nu\alpha\rho\sigma}k_{2}^{\mu})k_{1,\rho}k_{2,\sigma}+(k_{1}k_{2})\varepsilon^{\mu\nu\alpha\rho}(k_{2}-k_{1})_{\rho})\nonumber\\
+ & f_{2}(k_{1}+k_{2})^{\alpha}\varepsilon^{\mu\nu\rho\sigma}k_{1,\rho}k_{2,\sigma}+f_{3}%
(k_{1}+k_{2})^{\alpha}(g^{\mu\nu}(k_{1}k_{2})-k_{1}^{\nu}k_{2}^{\mu})\nonumber\\
+ & f_{4}((k_{1}^{2}g^{\mu\rho}-k_{1}^{\mu}k_{1}^{\rho})(g^{\nu\alpha}%
k_{2}^{\rho}-g^{\nu\rho}k_{2}^{\alpha})+(k_{2}^{2}g^{\nu\rho}-k_{2}^{\nu}%
k_{2}^{\rho})(g^{\mu\alpha}k_{1}^{\rho}-g^{\mu\rho}k_{1}^{\alpha})) \ .
\label{AGGStruct}
\end{align}
One can check that the eight operators in Eq.~(\ref{AGG3}) contribute to only these four $f_i$ form-factors. And once squared and summed over the polarization, the rate $\Gamma(A\rightarrow\gamma\gamma)$ does automatically vanish. 

Moving on to higher dimensions, the only new feature is the presence of operators that are invariant under a dark gauge symmetry, with for example, 
\begin{equation}
    \mathcal{L}^{\text{dim }8}_{A\gamma\gamma}\supset \frac{1}{\Lambda^4}F_{\mu\nu}\partial^{\lambda}\mathcal{F}_{\mu\nu}\partial^{\rho}\mathcal{F}_{\rho\lambda}\ ,\ \frac{1}{\Lambda^4}F_{\mu\nu}\partial^{\lambda}\mathcal{\tilde{F}}_{\mu\nu}\partial^{\rho}\mathcal{F}_{\rho\lambda}\ , \frac{1}{\Lambda^4}\varepsilon_{\mu\nu\rho\sigma}F^{\mu\lambda}\partial_{\lambda}\mathcal{F}^{\eta\nu}\partial^{\rho}\mathcal{F}_{\eta}^{\;\sigma}\label{AGG4}.
\end{equation}
If the first and second vanish through the photons EOM, the last cannot be trivially cancelled. Nevertheless, in terms of the tensor decomposition of Eq.~(\ref{AGGStruct}), all dimension-eight operators can do is to generate some $k^2_1$, $k^2_2$, and $(k_1+k_2)^2$ dependencies for the $f_i$ form factors, and $|\mathcal{M}(A\rightarrow\gamma\gamma)|^2$ still vanishes when squared and summed over polarization.

We now focus on the tensor embedding for the dark photon, and to set the stage, we search for the leading effective operator able to induce $B\rightarrow\gamma\gamma$. All the dimension-five candidates vanish by symmetry, and it is to be noticed that the trick used to couple the $C$ field to the Chern-Simons form is no longer permitted. Indeed, gauge invariance would be lost if $B$ is massive since under $\mathcal{A}_{\mu}\rightarrow \mathcal{A}_{\mu}+\partial_{\mu}\chi$,%
\begin{equation}
F_{\mu\nu\rho}^{B}\mathcal{A}^{\mu}\mathcal{F}^{\nu\rho}\rightarrow F_{\mu\nu\rho}^{B}\mathcal{A}^{\mu}\mathcal{F}^{\nu\rho}+F_{\mu\nu\rho}^{B}\partial^{\mu}\chi\mathcal{F}^{\nu\rho}=F_{\mu\nu\rho}^{B}\mathcal{A}^{\mu}\mathcal{F}^{\nu\rho}-\partial^{\mu}F_{\mu\nu\rho}^{B}\chi\mathcal{F}^{\nu\rho}\ ,
\label{BGG1}
\end{equation}
but $\partial^{\mu}F_{\mu\nu\rho}^{B}=m_{B}^{2}B_{\nu\rho}$ (the other term arising by partial integration vanishes by the QED Bianchi identity). Therefore, the first non-vanishing operators arise at the dimension-seven level, of which we can quote a few:
\begin{equation}
\mathcal{L}^{\text{dim }7}_{B\gamma\gamma} \supset \frac{1}{\Lambda^3} F_{\mu\nu\rho}^{B}\mathcal{F}^{\mu}_{\;\sigma}\partial^{\rho}\mathcal{F}^{\sigma\nu}\ ,\   \frac{1}{\Lambda^3}B_{\mu\nu}\partial^{\lambda}\mathcal{F}^{\mu\nu}\partial^{\rho}\mathcal{F}_{\rho\lambda}\ ,\ \frac{1}{\Lambda^3}\varepsilon_{\mu\nu\rho\sigma}B^{\mu\lambda}\partial_{\lambda}\mathcal{F}^{\eta\nu}\partial^{\rho}\mathcal{F}_{\eta}^{\;\sigma} \ , \ \text{etc} \ . 
\label{BGG2}
\end{equation}
Actually, to simplify this construction, we can notice that all the operators arising at this level can be obtained from the dimension-six or eight operators for the vector, \eqref{AGG3} and \eqref{AGG4}, using \eqref{BtoA} which either reduces or increases their mass dimension. Notice though that there is a subtlety as operators vanishing by the Lorenz condition are mapped onto operators vanishing upon imposing the Bianchi identity: 
\begin{equation}
\partial^{\mu}A_{\mu}\rightarrow \varepsilon_{\mu\nu\rho\sigma}\partial^{\mu}F_B^{\nu\rho\sigma}=0 \ \ , \ \ \ \partial^{\mu}B_{\mu\nu} \rightarrow \varepsilon_{\mu\nu\rho\sigma
}\partial^{\mu}F^{A,\rho\sigma} = 0 \ .
\end{equation}
This is irrelevant for our purpose of inducing $A/B \rightarrow \gamma\gamma$. Starting from this basis of operators, one could check that $|\mathcal{M}(B\rightarrow\gamma\gamma)|^2$ systematically cancels out after imposing the kinematics. 

More generally, a very neat demonstration can be done based on how duality relates squared amplitudes. At the core of these relationships are correspondences between the polarization sums for on-shell dual states, see Ref.~\cite{Plantier:2025hcm}. In the present case, every operator involving $F^B_{\mu\nu\rho}$ has a dimension-six counterpart in terms of $A_{\mu}$, which at the amplitude level translates as
\begin{equation}
\varepsilon_{\mu\nu\rho\sigma}q^{\nu}\epsilon^{\rho\sigma}(q) \rightarrow \Lambda \epsilon_{\mu}(q) \ .
\end{equation}
Similarly, any operator involving $B^{\mu\nu}$ has a dimension-eight counterpart involving $F_A^{\mu\nu}$, and their contributions at the amplitude level are related via
\begin{equation}
\ \ \epsilon_{\mu\nu}(q) \ \rightarrow \frac{1}{\Lambda} \varepsilon_{\mu\nu\rho\sigma}q^{\rho}\epsilon^{\sigma}(q) \ .
\end{equation}
Thus, the $\mathcal{M}(B\rightarrow\gamma\gamma)$ amplitude matches onto the tensor structure of Eq.~(\ref{AGGStruct}). In other words, though the expressions of the $f_i$ form-factors in terms of the operator coefficients is different, the total rate vanishes in both the $A$ and $B$ cases.

The contributions arising from fermion loops dutifully match onto this pattern of effective operators. With a tensor coupling to fermions, $B_{\mu\nu}\bar{\psi}\sigma^{\mu\nu}\psi$ and $B_{\mu\nu}\bar{\psi}\sigma^{\mu\nu}\gamma^{5}\psi$, the triangle amplitude is $TVV$ or $\tilde{T}VV$, which are both exactly zero by Furry's theorem. One could instead take an effective dimension-five coupling to fermions, $F_{\mu\nu\rho}^{B}\varepsilon^{\mu\nu\rho\sigma}(\bar{\psi}\gamma_{\sigma}\gamma^{5}\psi)$, in which case it is the $AVV$ triangle that matters. To preserve QED gauge invariance, it has to be in the Sutherland-Veltman configuration, Eq.~(\ref{AVVgen}) with $a=b=0$, leaving $T_{AVV}$ of $\mathcal{O}(m^{-2})$. These contributions then match onto the dimension-seven operators in Eq.~(\ref{BGG2}). Regarding the vector embedding, it similarly receives fermionic loop contributions from the $AVV$ triangle in the Sutherland-Veltman configuration, that is, when its leading term is of $\mathcal{O}(m^{-2})$, which maps onto the dimension-six operators of Eq.~(\ref{AGG3}). 

There are two possible way out of the Laudau-Yang situation. The first is rather trivial: if one of the photon is off-shell, then the decay is no longer forbidden. This is well known in the context of the gluon decay of the Z boson~\cite{Dicus:1985wx,vanderBij:1988ac, Lee:1988mz}, $Z\rightarrow ggg$, where there is a triangle contribution $Z\rightarrow gg$ followed by a $g \rightarrow gg$ vertex. Similarly, we expect $A/B \rightarrow \gamma(\gamma^{\ast}\rightarrow e^+e^-)$ to exist. Yet, with the operator structure established in the previous section, we expect $B\rightarrow\gamma\gamma^{\ast}$ to be significantly more suppressed as the corresponding operators are at best scaled by $\Lambda^{-3}$ against $\Lambda^{-2}$ for the vector.

The second way to evade Landau-Yang is more delicate. Actually, one should first remember the condition allowing us to compute $Z\rightarrow gg^*$ or $Z\rightarrow \gamma\gamma^*$ in the first place, which is the cancellation of the SM anomalies in the corresponding triangles. To pinpoint where this mechanism enters, let us take a Stueckelberg representation for the massive vector field
\begin{equation}
A_{\mu} \rightarrow A_{\mu}-\frac{1}{m_{A}}\partial_{\mu}\phi\ . \label{AGG5}%
\end{equation}
Gauge invariance is ensured by the coherent transformations $A_{\mu}\rightarrow A_{\mu}+\partial_{\mu}\chi$ together with $\phi\rightarrow\phi+m_A\chi$. Consider now the $AVV$ triangle contribution, $\mathcal{M}(A\rightarrow\gamma
\gamma)=T_{AVV}^{\alpha\mu\nu}\varepsilon^{\alpha}$, with $\varepsilon^{\alpha}$ the polarization vector of the massive vector, while those of the photons are kept understood for clarity. In the unitary gauge $\chi=-\phi/m_{A}$, one gets
\begin{equation}
\sum_{pol}|\mathcal{M}(A(k)\rightarrow\gamma\gamma)|^{2}=\left(
-g_{\alpha\beta}+\frac{1}{m_{A}^{2}}k_{\alpha}k_{\beta}\right)  T_{AVV}^{\alpha\mu\nu}T_{AVV}^{\ast\beta\rho\sigma}\ .
\label{AggProc}
\end{equation}
The Landau-Yang theorem is then active, and when the summation over the photon polarization is performed, the total rate vanishes. 
By contrast, in the representation of Eq.~(\ref{AGG5}), which is essentially the Feynman gauge~\cite{Plantier:2025hcm}, the $\phi$ coupling to fermions generates a pseudoscalar coupling,
$\partial_{\mu}\phi\bar{\psi}\gamma^{\mu}\gamma^{5}\psi\rightarrow2im\phi
\bar{\psi}\gamma^{5}\psi$, and the total rate is rather
\begin{equation}
\sum_{pol}|\mathcal{M}(A(k)\rightarrow\gamma\gamma)|^{2}+|\mathcal{M}%
(\phi(k)\rightarrow\gamma\gamma)|^{2}=-g_{\alpha\beta}T_{AVV}^{\alpha\mu\nu
}T_{AVV}^{\ast\beta\rho\sigma}+\left(  \frac{2im}{m_{A}}T_{PVV}^{\mu\nu
}\right)  \left(  \frac{2im}{m_{A}}T_{PVV}^{\ast\rho\sigma}\right)  \ .
\end{equation}
Both results are compatible provided $k_{\alpha}T_{AVV}^{\alpha\mu\nu}=2imT_{PVV}^{\mu\nu}$, which is the classical Ward identity. For the $Z$ boson, this identity holds because anomalies cancel, the $Z\rightarrow \gamma\gamma$ amplitude is gauge invariant, and thus always vanishes by virtue of the Landau-Yang theorem.

If the triangle anomaly does not cancel for the dark vector, it breaks the axial gauge invariance and the classical Ward identity fails. In the Stueckelberg representation, the $A\rightarrow\gamma\gamma$ and $\phi\rightarrow\gamma\gamma$ amplitudes no longer add to the same result in all gauges. In other words, the two-photon production rate depends on how much of the $\phi$ scalar is transmuted into the $\left\vert J,J_{3}\right\rangle =\left\vert 1,0\right\rangle$ polarization state of the massive vector, and how much is left as a true scalar $\left\vert 0,0\right\rangle $ state. The Landau-Yang theorem only holds for $J=1$, so if there is a $\left\vert 0,0\right\rangle $ left-over, two-photon productions will still arise from $\phi\rightarrow\gamma\gamma$. 

Let us stress that this is not truly a violation of the Landau-Yang theorem. Being kinematical, it cannot be escaped. Rather, one should account for an additional contribution originating from the $\partial_{\mu}\phi$ anomalous coupling to the Chern-Simons form, giving back the $\phi\mathcal{F}_{\mu\nu}\mathcal{\tilde{F}}^{\mu\nu}$ effective coupling discussed in the previous section, see Eq.~(\ref{phiFF}). The only difference is that here, the scale of those effective interactions is set by the dark photon mass. If the dark field is light, then the rate quickly increases. This is a typical feature for dark photon models, which always predicts poles in $1/m_{A}$ when the dark photon couplings to SM particles break the dark photon gauge invariance (see e.g. Refs.~\cite{Kamenik:2011vy, Kamenik:2012hn}).

A similar situation has not to be feared for a 2-form. Its Stueckelberg decomposition follows
\begin{equation}
B_{\mu\nu} \rightarrow B_{\mu\nu}-\frac{1}{m_{B}}F_{\mu\nu}\ ,\label{BGG5}%
\end{equation}
where $F_{\mu\nu}$ is the field strength of a vector field. This structure is compatible with the expected degrees of freedom of a $B$ field~\cite{Plantier:2025hcm}, which is dual to scalar field when massless, but to a vector field when massive. The point is that this time, the scalar part, $B_{\mu\nu}$, cannot contribute at leading order since the $TVV$ and $\tilde{T}VV$ triangle amplitudes are null. To be sensitive to the $AVV$ triangles requires to use the $F_{\mu\nu\rho}^{B}\varepsilon^{\mu\nu\rho\sigma}(\bar{\psi}\gamma_{\sigma}\gamma^{5}\psi)$, that is unaffected by the $F_{\mu\nu}$ component. This also means that the $F_{\mu\nu}$ part of Eq.~(\ref{BGG5}) cannot contribute anomalously. This is in direct correspondence with the absence of gauge invariant operators mediating $A\rightarrow \gamma\gamma$ before dimension eight. All in all, adopting a Stueckelberg form for the $B$ embedding, we do not expect any associated anomalous two-photon production. This is a stark breaking of duality. Fundamentally, though the massive $A$ and $B$ fields are dual, their scalar components are not because a massive scalar is dual to a massive 3-form field, as discussed in the previous section.

In conclusion, we expect the Landau-Yang theorem to be particularly effective for the $B$ embedding. Even allowing for one of the photons to be off-shell, $B\rightarrow\gamma\gamma^{\ast}$ arises from higher order operators compared to $A\rightarrow\gamma\gamma^{\ast}$. This situation is significantly reinforced in the presence of anomalies, with even the possibility to produce two on-shell photons from the scalar field associated to the generation of the dark vector mass, a phenomenon that has no counterpart for the 2-form representation.

\subsection{The dark spin-one pair productions : $\Phi_{SM}\rightarrow AA/BB$}

As a final case, we want to investigate the decay of a SM scalar or pseudoscalar particle in two dark spin-one particles. As before, we have to distinguish the cases in which $\Phi_{SM}$ is a scalar or a pseudoscalar, directly or derivatively coupled to the dark states.

If no dark gauge invariance is imposed, the dominant operators are simply the dimension-three%
\begin{equation}
\Lambda\Phi_{SM}A_{\mu}A^{\mu}\leftrightarrow\Lambda%
\Phi_{SM}B_{\mu\nu}B^{\mu\nu},\ \Lambda\Phi_{SM}B_{\mu\nu}\tilde
{B}^{\mu\nu}.\label{PhiVV1}%
\end{equation}
Both would lead to total rates of similar size, though the differential rate of $\Phi_{1,SM}\rightarrow\Phi_{2,SM}AA$ and $\Phi_{1,SM}\rightarrow\Phi_{2,SM}BB$ would be very different, see Ref.~\cite{Plantier:2025hcm}. It is to be noted also that because these operators break the dark gauge invariance, the longitudinal polarization states generate $m_{\Phi}^{2}/m_{A,B}^{2}$ and $m_{\Phi}^{4}/m_{A,B}^{4}$ terms in the total rates. This perfectly illustrates the comment made in the previous section after Eq.~(\ref{AggProc}).

The status of these operators is nevertheless quite different once fermion loops are considered. First, there are situations in which the loops do not contribute to the dimension-three operators. This arises if $\Phi$ is a pseudoscalar, because $\Phi\rightarrow AA$ involves the finite $PVV$ or $PAA$ triangles, both matching onto the dimension-five $\Phi\mathcal{F}_{\mu\nu}\mathcal{\tilde{F}}^{\mu\nu}$ operator. Similarly, if $\Phi$ is a scalar and $A$ has only vector couplings to fermions, then the $\Phi\rightarrow AA$ loop involves the finite $SVV$ triangle that matches onto $\Phi\mathcal{F}_{\mu\nu}\mathcal{F}^{\mu\nu}$ (see Eq.~(\ref{PVVPAA})). In these three cases, the effective coupling ends up being invariant under the dark gauge symmetry. Fundamentally, the reason for that comes from the special nature of the coupling to fermions in the vector case, $A_{\mu}\bar{\psi}\gamma^{\mu}\psi$, which via a change of basis is equivalent to the gauge invariant kinetic mixing $F_{\mu\nu}^{A}\mathcal{F}^{\mu\nu}$. For the $PAA$ loop, its finiteness is rather accidental since the $\bar{\psi}\gamma^{\mu}\gamma^{5}\psi$ current is not conserved. It rather rests on the similarity between the $PAA$ and $PVV$ triangle amplitudes, which share the same UV structure.

In these three specific cases, if one could argue against the presence of the couplings in Eq.~(\ref{PhiVV1}), the final $\Phi_{SM}\rightarrow AA$ rate would be rather suppressed, arising only from dimension-five operators. By contrast, for all the other cases, the dimension-three operators have to be present. Indeed, there is then no hidden dark gauge invariance and the fermion loop diverges, whether it is via the $SAA$ or tensor triangles. As explained in the first section, these divergences simply renormalize the non-gauge invariant couplings in Eq.~(\ref{PhiVV1}). In those situations, the counterterms would clearly be dominant and lead to similar rates for the $\Phi_{SM}\rightarrow AA/BB$ processes.

The only case in which $\Phi_{SM}\rightarrow AA$ could be structurally larger than $\Phi_{SM}\rightarrow BB$ is if the dark gauge invariance is imposed. At first sight we have the similar dimension-five couplings%
\begin{equation}
\frac{1}{\Lambda}\Phi_{SM}F_{\mu\nu}^{A}F^{A,\mu\nu}\ ,\ \frac{1}{\Lambda}%
\Phi_{SM}F_{\mu\nu}^{A}\tilde{F}^{A,\mu\nu}\leftrightarrow\frac{1}{\Lambda
}\Phi_{SM}F_{\mu\nu\rho}^{B}F^{B,\mu\nu\rho}\ .\label{PhiVV2}%
\end{equation}
Yet, if those are to be generated by fermion loops, only that for $A$ could arise, via the $PVV$, $PAA$, or $SVV$ triangles described above. On the contrary, for $B$, we would have first to restrict its coupling to fermions to $F_{\mu\nu\rho}^{B}\varepsilon^{\mu\nu\rho\sigma}(\bar{\psi}\gamma_{\sigma}\psi)$ and $F_{\mu\nu\rho}^{B}\varepsilon^{\mu\nu\rho\sigma}(\bar{\psi}\gamma_{\sigma}\gamma^{5}\psi)$, and the final loop would match onto dimension-seven operators.

Finally, in case $\Phi_{SM}$ has to be derivatively coupled, the leading operators are then again those in Eq.~(\ref{PhiVV2}), and the $\Phi_{SM}\rightarrow AA$ and $\Phi_{SM}\rightarrow BB$ rates are expected to be of similar size. This remains true if these operators are induced by fermion loops with an effective $\partial_{\mu}\Phi_{SM}\bar{\psi}\gamma^{\mu}\psi$ or $\partial_{\mu}\Phi_{SM}\bar{\psi}\gamma^{\mu}\gamma^{5}\psi$ coupling, since the $AVV$, $AAA$, and $ATT$, $A\tilde{T}\tilde{T}$ triangles all match onto these operators. Notice though that in the tensor case, the loop actually diverges, so the operators in Eq.~(\ref{PhiVV2}) have in any case to be present from the start, and the value of their Wilson coefficient cannot be expected to be loop suppressed.

In conclusion, we find that the expected relative size of $\Gamma(\Phi_{SM}\rightarrow AA)$ and $\Gamma(\Phi_{SM}\rightarrow BB)$ is strongly dependent on the underlying assumptions. In most cases, both are expected to scale similarly with $\Lambda$, though in specific scenarios, one could arrange for either one to be significantly suppressed.

\section*{Conclusion}

An immediate way to see the phenomenological interest of dark higher-forms is through the peculiar nature of their dominant couplings with fermions: the 2-form field couples to a (pseudo)tensor current and the 3-form couples to a vector or an axial current through a Levi-Civita tensor. The fact that these interactions are renormalizable opens the gate to the consideration of loop processes containing (pseudo)tensor or anti-symmetric vertices, that were previously considered irrelevant in the context of the SM. We focused in this article on two simple processes: $\Phi_{SM}\rightarrow XX$ and $X\rightarrow \gamma\gamma$ with $X=\phi,A,B,C$, in order to see how they could be affected by fermionic triangle loop contributions. This required us to calculate explicitly the triangles containing (pseudo)tensor couplings. Our main findings are
\begin{itemize}
    \item A triangle containing a (pseudo)tensorial vertex cannot be anomalous. We can see this classifying tensorial triangles in three categories: First, the ones containing an even number of axial vertices are all cancelled out by Furry theorem. Among the few surviving triangles, a second category contains the amplitudes that have the fermion-mass as an overall factor, as for example the $AVT$ triangle. The fact that these amplitudes vanish identically in the massless limit is enough to prove the absence of anomalies. The third category contains the triangle amplitudes that exhibit a UV-pole. This situation systematically arises when a renormalizable operator actually exist. As a consequence, any \textit{would-be} anomaly coming from the axial-tensorial triangle could be absorbed into some renormalizable couplings, making it effectively harmless. For comparison, the usual $AVV$ triangle sits somewhat between the second and third categories discussed above. It has a surviving mass-independent term that no renormalizable operators can absorb.
    \item For the pair production of scalar dark matter particles, duality remains effective, with $\Phi_{SM}\rightarrow\phi\phi$ or $\Phi_{SM}\rightarrow CC$ processes being of the same order in the scale $\Lambda$. Notice though that differential rates would be significantly different, see Ref.~\cite{Plantier:2025hcm}. Also, it should be stressed that while $\Phi_{SM}\rightarrow\phi\phi$ becomes automatically forbidden when $\Phi_{SM}$ is pseudoscalar, it requires the Lorenz condition $\partial^{\mu}C_{\mu\nu\rho}=0$ for $\Phi_{SM}\rightarrow CC$. 
    \item The Landau-Yang theorem is proven to hold for both a vector and a 2-form field, but we find there are still very significant differences. First, $A\rightarrow\gamma\gamma$ receives contributions from $AVV$ that only subsist if one of the photons is off-shell. However, $B\rightarrow\gamma\gamma$ cannot receive any loop contribution when starting only from the renormalizable $B_{\mu\nu}\bar{\psi}\sigma^{\mu\nu}\psi$ coupling, not even off shell as $TVV=\tilde{T}VV=0$ by Furry's theorem. Only higher-order contributions starting from an effective $F^B_{\mu\nu\rho}\varepsilon^{\mu\nu\rho\sigma}\bar{\psi}\gamma_{\sigma}\gamma^5\psi$ coupling could arise. Second, the UV mechanisms making the $A$ or $B$ massive are not affected in the same way by anomalies. In a Higgs or Stueckelberg representation, the $A$ field would be accompanied by a scalar field which, in the presence of the $AVV$ anomaly, could decay into two real photons. By contrast, the $B$ field can be made massive by the presence of a vector field, that cannot decay into on-shell photons. 
    \item One striking result is the fact that a non-null, non-suppressed $C\rightarrow \gamma\gamma$ decay at loop level can arise by an explicit violation of the vector Ward identity in the $AVV$ triangle. This is viable even if the vector currents stand for photons or gluons because the $C$ field actually couples to the Chern-Simons form $G^{\mu\nu\rho}$ via the renormalizable operator $C_{\mu\nu\rho}G^{\mu\nu\rho}$, which is gauge invariant thanks to the Lorenz condition $\partial^{\mu}C_{\mu\nu\rho}=0$. In some sense, the $AVV$ anomaly has to be moved to the vector current since it is the presence of the gauge fields that makes the axial current anomalous. By contrast, $\phi\rightarrow \gamma\gamma$ proceeds from the dimension-five coupling $(\partial_{\sigma}\phi) \varepsilon^{\mu\nu\rho\sigma}G_{\mu\nu\rho}$, so we expect $\Gamma (C\rightarrow\gamma\gamma)\gg\Gamma (\phi\rightarrow\gamma\gamma)$.
   \item Regarding the decay of $\Phi_{SM}$ in two spin-one particles, the predominance of the vector or the 2-form channel depends entirely on the specificity of the underlying scenarios. For example, if one imposes the dark gauge invariance, the decay receives only loop contributions which are immediately gauge invariant for the vector. In this context, $\Gamma(\Phi_{SM}\rightarrow AA)$ largely dominates. However, if one invokes specific parity arguments, the existence of the peculiar dimension-three pseudoscalar coupling $\Lambda\Phi_{SM}B_{\mu\nu}\tilde{B}^{\mu\nu}$ could be responsible for the preponderance of $\Gamma(\Phi_{SM}\rightarrow BB)$.   
   
\end{itemize}
Higher-forms are once again proven to provide a compelling alternative framework. If they cannot overstep the physical impossibilities related to the nature of the particles they embody (e.g. $B$ has to obey the Landau-Yang theorem), the fact that duality shuffles parity, gauge-invariance and mass-scaling still drives to important phenomenological distinctions both at the level of effective operators and triangle loops. Among the theoretical outcomes of this paper, the peculiar $C_{\mu\nu\rho}G^{\mu\nu\rho}$ coupling, its emergence at loop level, and its indirect gauge-invariance would deserve further investigations, especially in the context of QCD and its vacuum structure.
\subsubsection*{Acknowledgments:}

This research is supported by the IN2P3 Master project \textquotedblleft
Axions from Particle Physics to Cosmology\textquotedblright, and from the
French National Research Agency (ANR) in the framework of the
\textquotedblleft GrAHal\textquotedblright project no. ANR-22-CE31-0025.

\bibliographystyle{BiblioStyle}
\bibliography{references}

\end{document}